\documentclass[12pt]{article}
\usepackage{amsmath,amssymb,epsfig, bm,color,psfrag,pstool,subcaption}
\usepackage{feynmf,graphicx}

\allowdisplaybreaks

\newcommand{\nl}{\nonumber \\ }
\def\kslash{\rlap{\hspace{-0.005cm}/}{k}}
\def\lslash{\rlap{\hspace{-0.03cm}/}{l}}
\def\qslash{\rlap{\hspace{-0.01cm}/}{q}}
\def\Dslash{\rlap{\hspace{0.07cm}/}{D}}
\newcommand{\q}{f}
\newcommand{\LL}{q}
\newcommand{\order}{{\cal O}}
\newcommand{\sym}{S}
\newcommand{\be}{\begin{equation}}  
\newcommand{\ee}{\end{equation}} 

\makeatletter
\newcommand{\vast}{\bBigg@{4}}
\newcommand{\Vast}{\bBigg@{5}}
\makeatother

\begin{document}
\begin{fmffile}{fmf_W1}

\fmfcmd{%
  vardef cross_bar (expr p, len, ang) =
  ((-len/2,0)--(0,0))
  rotated (ang + angle direction length(p) of p) shifted point length(p) of p
  enddef;
  style_def crossed expr p =
  cdraw p;
  ccutdraw cross_bar (p, 2mm, 30); ccutdraw cross_bar (p, 2mm, -30)
  enddef;}
\fmfcmd{%
  vardef port (expr t, p) =
  (direction t of p rotated 90)
  / abs (direction t of p)
  enddef;}
\fmfcmd{%
  vardef portpath (expr a, b, p) =
  save l; numeric l; l = length p;
  for t=0 step 0.1 until l+0.05:
  if t>0: .. fi point t of p
  shifted ((a+b*sind(180t/l))*port(t,p))
  endfor
  if cycle p: .. cycle fi
  enddef;}
\fmfcmd{%
  style_def brown_muck expr p =
  shadedraw(portpath(thick/2,2thick,p)
  ..reverse(portpath(-thick/2,-2thick,p))
  ..cycle)
  enddef;}

\fmfset{arrow_len}{3mm}
\fmfset{arrow_ang}{12}
\fmfset{wiggly_len}{3mm}
\fmfset{wiggly_slope}{75}
\fmfset{curly_len}{2mm}

\begin{titlepage}

\begin{flushright}
FERMILAB-PUB-16-465-T \\
WSU-HEP-1607 \\
November 29, 2016
\end{flushright}

\vspace{0.7cm}
\begin{center}
\Large\bf 
Nucleon spin-averaged forward virtual Compton tensor at large $Q^2$
\end{center}

\vspace{0.8cm}
\begin{center}
{\sc  Richard J. Hill$^{(a)}$ and Gil Paz$^{(b)}$  } \\
\vspace{0.4cm}
{\it 
  $^{(a)}$
  Perimeter Institute for Theoretical Physics, Waterloo, ON, N2L 2Y5 Canada, \\
  TRIUMF, 4004 Wesbrook Mall, Vancouver, BC, V6T 2A3 Canada, \\
  Fermi National Accelerator Laboratory, Batavia, Illinois 60510, USA, and \\
  The University of Chicago, Chicago, Illinois, 60637, USA
}\\
\vspace{0.7cm}
{\it 
$^{(b)}$ 
Department of Physics and Astronomy \\
Wayne State University, Detroit, Michigan 48201, USA}
\end{center}
\vspace{1.0cm}

\begin{abstract}
\vspace{0.2cm}
\noindent  

The nucleon spin-averaged forward virtual Compton tensor
determines important physical quantities such as
electromagnetically-induced mass differences of nucleons, and
two-photon exchange contributions in hydrogen spectroscopy.  It
depends on two kinematic variables: $\nu$, the virtual photon energy
in the proton rest frame, and $Q^2$, the photon's invariant
four-momentum squared.  Using the operator product expansion, we
calculate the tensor's large-$Q^2$ behavior for $\nu=0$, including for
the first time the full spin-2 contribution and correcting a previous
result in the literature for the spin-0 contribution.  Implications for the proton
radius puzzle are discussed. 

\vfil
\end{abstract}
\end{titlepage}

\section{Introduction}

\begin{figure}[tb]
  \begin{center}
    \includegraphics[height=0.15\textwidth]{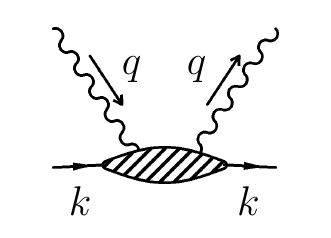}
    \vspace{5mm}
    \caption{Forward scattering of virtual photon of momentum $q^\mu$ on hadron of momentum $k^\mu$.  \label{fig:T}
    }
  \end{center}
\end{figure}

The forward Compton amplitude describes virtual photon scattering
on a hadron, depicted in Fig.~\ref{fig:T}, and encodes important
structure information.  For example, the spin-averaged proton forward
Compton amplitude determines two-photon exchange (TPE) contributions
to hydrogen spectroscopic transitions.  These contributions dominate
the uncertainty of the proton charge radius measured in muonic
hydrogen~\cite{Pohl:2010zza} and have been the subject of recent intense scrutiny and
controversy~\cite{Hill:2011wy,Birse:2012eb,Antognini:2013jkc,Peset:2014jxa}.
The difference between spin-averaged proton and neutron forward
Compton amplitudes determines the electromagnetic contribution to the
proton-neutron mass splitting, also the subject of recent
controversy~\cite{Cottingham:1963zz,Collins:1978hi,WalkerLoud:2012bg,Erben:2014hza,Gasser:2015dwa}.

The tensor representation of the spin-averaged Compton amplitude
(cf. Eq.~(\ref{eq:hadten}) below) involves two scalar functions,
$W_1(\nu,Q^2)$ and $W_2(\nu,Q^2)$.  The imaginary parts of these
functions are determined by (elastic and inelastic) electron
scattering data.   $W_2(\nu,Q^2)$ can be reconstructed from its
imaginary part using  a dispersion relation in $\nu$.  However,
$W_1(\nu,Q^2)$ requires a subtraction in the dispersion relation.
Uncertainties related to the subtraction function for the proton,
$W_1^p(0,Q^2)$, dominate the
error budget for the muonic hydrogen Lamb shift.  Similarly,
uncertainties in the difference of subtraction functions for proton and
neutron, $W_1^{p-n}(0,Q^2)$, dominate the error budget for the  
proton-neutron electromagnetic self energy.  It is important to
constrain this subtraction function.%
\footnote{
  The need for subtraction is readily demonstrated: $W_1(\nu,0) = -2 + \order(\nu^2)$
  at low $\nu$ (see e.g. Ref.~\cite{Hill:2012rh}), in contradiction with the prediction from
  an unsubtracted dispersion relation,
  $W_1(\nu,0) = (1/\pi) \int_{\nu_{\rm cut}^2}^\infty d\nu^{\prime 2} {\rm Im}W_1(\nu^\prime,0)/(\nu^{\prime 2} - \nu^2)$,
  which is manifestly positive using the optical theorem for ${\rm Im}W_1(\nu^\prime,0)$ (see e.g. Ref.~\cite{Hill:2011wy}). 
  Reference~\cite{Gasser:2015dwa} argues that after subtraction of a term
  motivated by Reggeon dominance, the remaining structure functions
  satisfy unsubtracted dispersion relations; however, a reliable evaluation
  of such a subtraction has not been presented. 
}

While not accessible directly from experimental data, we can constrain
$W_1(0,Q^2)$ in the small- and large-$Q^2$ limits. For small $Q^2$ one
can  calculate~\cite{Hill:2011wy} $W_1(0,Q^2)$ in terms of the Wilson
coefficients of Non Relativistic QED (NRQED)~\cite{Caswell:1985ui,Hill:2012rh}.
These coefficients are
in turn determined by low-energy experimental measurements, or by
nonperturbative QCD computations using lattice QCD or chiral
perturbation theory.  The Operator Product Expansion (OPE) may be used
to show that $W_1(0,Q^2)\sim1/Q^2$ for large $Q^2$
\cite{Collins:1978hi, Hill:2011wy}.  In this paper we perform the
explicit computation of the $1/Q^2$ coefficient in the OPE
determination of $W_1(0,Q^2)$.

There are four types of operator that contribute to the coefficient of $1/Q^2$
in the OPE expression for $W_1(0,Q^2)$: spin-0 quark, spin-0 gluon, spin-2 quark
and spin-2 gluon.
Previous analyses of $W_1(0,Q^2)$ are based on the work of
Collins from 1979~\cite{Collins:1978hi}, which was concerned with the
renormalization of the Cottingham formula for electromagnetic mass splittings,
and considered only the spin-0 operators.
In this paper we calculate the Wilson coefficients for both types of operator.
We correct an error in the expression for
the spin-0 contribution in Ref.~\cite{Collins:1978hi}. 
Applications to muonic hydrogen require
inclusion of the spin-2 operators.  Since these were not included
before, existing applications are incomplete; in fact, we show that the spin-2
contribution is numerically dominant.    

The remainder of the paper is structured as follows. In
Sec.~\ref{Analytical} we perform the analytic calculations of the
Wilson coefficients.  In particular we perform the one-loop
calculations needed to extract the Wilson coefficients of gluon
operators. In Sec.~\ref{Numerical} we evaluate the relevant local
nucleon matrix elements.  Focusing on the proton case, we
numerically evaluate $W_1(0,Q^2)$ at large $Q^2$.  In
Sec.~\ref{Applications} we discuss the implications of our
results for TPE contributions to the muonic hydrogen Lamb shift.
In Sec.~\ref{Conclusions} we present our conclusions.
 
\section{Wilson coefficients} \label{Analytical}

Here we introduce notation for the Compton
amplitude and determine the relevant operators in the OPE.
We then perform matching onto quark and gluon operators. 

\subsection{Preliminaries\label{sec:prelim}}

The nucleon spin-averaged forward virtual Compton scattering tensor is
defined as (cf. Fig.~\ref{fig:T})
\begin{equation}
\label{eq:hadten}
W^{N\,\mu\nu}_{\rm \sym}(\nu,Q^2)\equiv \frac{i}{2}\sum_s\int
d^4x\,e^{iq\cdot x} \langle N(\bm{k},s)| T\{ J_{\rm e.m.}^\mu(x) J_{\rm e.m.}^\nu(0) \} | N(\bm{k},s) \rangle\,,
\end{equation}
where $\bm{k}$ is the nucleon three-momentum and $s$ its spin, 
$J^{\mu}_{\rm e.m.}$ is the electromagnetic current, and 
the subscript ${\rm \sym}$ denotes symmetrization in $\mu$ and $\nu$.%
\footnote{Symmetrization in $\mu$ and $\nu$ can be shown equivalent to spin averaging:
  if $W^{N\,\mu\nu}$ is expanded in a tensor basis, terms that are odd under $\mu \leftrightarrow \nu$
  are linear in the nucleon spin vector, $S^\mu \propto \bar{u}_N(p,s) \gamma^\mu \gamma_5 u_N(p,s)$,  and
  vanish under spin averaging;
  terms symmetric under $\mu \leftrightarrow \nu$ are independent of $S^\mu$ as displayed in
  Eq.~(\ref{eq:decompose}). 
}
Here $N=p$ or $N=n$ denotes
a proton or neutron.   Also,
$\nu=2 k\cdot q$ and $Q^2=-q^2$.  Notice that some authors refer to
this quantity as $T^{\mu\nu}$ \cite{Collins:1978hi,
  WalkerLoud:2012bg}.

Using current conservation, and invariance of electromagnetic
interactions under parity  and time-reversal,  $W_{\rm \sym}^{N\, \mu\nu}$ can be
expressed as 
\begin{equation}\label{eq:decompose}
W_{\rm \sym}^{N\,\mu\nu}=\left( - g^{\mu\nu} + \frac{q^\mu q^\nu}{ q^2} \right) W_1^N(\nu,Q^2)
+ 
\left( k^\mu - \frac{k\cdot q \,q^\mu }{q^2} \right) 
\left( k^\nu - \frac{k\cdot q \, q^\nu} {q^2} \right) W_2^N(\nu,Q^2) \,. 
\end{equation}
We will use relativistic normalization of states, $\bar{u}_N(k) u_N(k) = 2m_N$, for
nucleon spinors throughout. 
The imaginary parts $W_i^N$ are related to physical
cross sections. Inserting a complete set of states into
(\ref{eq:hadten}), 
\begin{align}\label{eq:ImW}
2\, {\rm Im}\, W_{\rm \sym}^{N\,\mu\nu}=
\frac12\sum_s 
\rlap{\hspace{0.15cm}$\displaystyle\int$}\sum_{\,X} 
\bigg[&\langle N(\bm{k},s)|J^\mu|X\rangle\langle X|J^\nu| N(\bm{k},s) \rangle
  (2\pi)^4\delta^4(q-p_X+k)\nl+&\langle N(\bm{k},s)|J^\nu|X\rangle\langle X|J^\mu| N(\bm{k},s) \rangle(2\pi)^4\delta^4(q+p_X-k)\bigg] \,,
\end{align} 
where $p_X$ is the momentum of the state $X$. We can now perform a
separation between nucleon and non-nucleon (i.e., excited) states. The nucleon
contributions  to ${\rm Im}\, W_i^N$ can be expressed in terms  of
nucleon form factors.  Contributions from other final states can be
expressed in terms of inelastic structure functions.  Using dispersion
relations,  $W_2^N$ can be reconstructed from its imaginary part, but $W_1^N$
requires a subtraction in order to have a convergent dispersion
relation. Thus 
\begin{eqnarray}\label{dispersion}
  W_1^N(\nu,Q^2) &=& W_1^N(0,Q^2)+
  {\nu^2\over \pi} \int_{0}^\infty { d\nu^{\prime 2}} 
\,{ {\rm Im}W_1^N(\nu^\prime, Q^2) \over \nu^{\prime 2} (\nu^{\prime 2} - \nu^2) }  \,,
\nonumber\\\nonumber\\
W_2^N(\nu,Q^2) &=&
{1\over \pi} \int_{0}^\infty { d\nu^{\prime 2} }
\,{ {\rm Im}W_2^N(\nu^\prime, Q^2) \over \nu^{\prime 2} - \nu^2}
\,.
\end{eqnarray} 

Since $W_1^N(0,Q^2)$ is not directly related to measured quantities it
is a major source of uncertainty for quantities like the TPE
contribution to the Lamb shift in muonic hydrogen~\cite{Hill:2011wy}
and the isovector nucleon electromagnetic
self-energy~\cite{WalkerLoud:2012bg}. As discussed in the introduction,
we can constrain $W_1^N(0,Q^2)$ in the small-$Q^2$ and large-$Q^2$ limits.
The small-$Q^2$ behavior was discussed in Ref.~\cite{Hill:2011wy}. Here we will
calculate the large $Q^2$ behavior. 

We will determine the OPE for the operator, 
\begin{align}
  T_{\rm \sym}^{\mu\nu}(q)\equiv i\int d^4x\,e^{iq\cdot x}  T\{ J_{\rm e.m.}^{\{\mu}(x) J_{\rm e.m.}^{\nu\}}(0) \}
  \,,
\end{align}
where curly braces around indices denote symmetrization,
i.e., $A^{\{\alpha}B^{\beta\}}=(A^\alpha B^\beta+ B^\alpha A^\beta)/2$ for four-vectors $A^\mu$ and $B^\mu$. 
Note that the Compton tensor is obtained by taking the matrix element,
$W_{\rm \sym}^{N\,\mu\nu}(\nu,Q^2)= \langle N(\bm{k},s)| T_{\rm \sym}^{\mu\nu}| N(\bm{k},s) \rangle$.
For the OPE evaluation we match onto the lowest
dimension QCD operators, which begin at dimension four. There are
four%
\footnote{As we will see below, there is another possible
  structure $O_{\rm \sym}^{(r)\mu\alpha\nu\beta}$.
  The matrix element of the corresponding contribution to $T^{\mu\nu}_S$
  between proton states vanishes, so we do not include it in Eq.~(\ref{operator_basis}).
  See Appendix~\ref{sec:appendix}.
}
relevant operator types
\begin{eqnarray}\label{operator_basis}
O_{\q}^{(0)}&=&m_{\q}\bar\q \q,\nonumber\\
O_g^{(0)}&=&G^A_{\alpha\beta}G^{A\alpha\beta},\nonumber\\
O_{\q}^{(2)\alpha\beta}&=&  \bar \q\left( iD^{ \{\alpha} \gamma^{\beta\}} -\frac{1}{d}\,i\Dslash g^{\alpha\beta}\right)\q,\nonumber\\
O_g^{(2)\alpha\beta}&=&-G^{A\alpha\lambda}G^{A\beta}_{\quad\lambda}+\dfrac{1}{d} g^{\alpha\beta}G^A_{\rho\sigma}G^{A\rho\sigma},
\end{eqnarray}
where $d=4-2\epsilon$ is the spacetime dimension, and $\q=u,d,s,...$ runs over active quark flavors.
The superscript label denotes operator spin.  The spin-2 operators are traceless,
i.e. they satisfy $g_{\alpha\beta}O_{\q,g}^{(2)\alpha\beta}=0$.

Current conservation implies $q_\mu T^{\mu\nu}=q_\nu T^{\mu\nu}=0$ up to operators whose physical
matrix elements vanish.
This implies that the general form%
\footnote{A fourth possible term $q_\alpha q_\beta O_{\rm \sym}^{(r)\mu\alpha\nu\beta}$
  is omitted since its matrix elements between spin-averaged proton states is zero.}
of $T_{\rm \sym}^{\mu\nu}$ may be written
\begin{eqnarray}\label{eq:Tsymm}
  T_{\rm \sym}^{\mu\nu}&=&\dfrac1{q^2}\left(-g^{\mu\nu}+\dfrac{q^\mu q^\nu}{q^2}\right)
  \left(\sum_{\q} c_{1\q}O_{\q}^{(0)}+c_{1g}O_g^{(0)}\right) \nonumber\\
  &+&\dfrac1{q^2}\left(-g^{\mu\nu}+\dfrac{q^\mu q^\nu}{q^2}\right)
  \dfrac{q_\alpha q_\beta}{q^2}\left(\sum_{\q} c_{2\q}O_{\q}^{(2)\alpha\beta}+c_{2g}O_g^{(2)\alpha\beta}\right) \nonumber\\
  &+&\dfrac1{q^2}\left(-g^{\mu}_{\alpha}+\dfrac{q^\mu q_\alpha}{q^2}\right)
  \left(-g_{\beta}^{\nu}+\dfrac{q_\beta q^\nu}{q^2}\right)
  \left(\sum_{\q} c_{3\q}O_{\q}^{(2)\alpha\beta}+c_{3g}O_g^{(2)\alpha\beta}\right)+\order\left(\frac1{q^4}\right).\qquad
\end{eqnarray}
Using the matrix elements (\ref{eq:scalar}) and (\ref{eq:tensor}) below, we see that $c_1$ and
$c_2$ will contribute to $W_1$ and $c_3$ to $W_1$ and $W_2$.  
Let us proceed to match onto quark and gluon operators at leading order
in QCD perturbation theory.

\subsection{Quark operators}

\begin{figure}[tb]
  \begin{center}
    \includegraphics[height=0.15\textwidth]{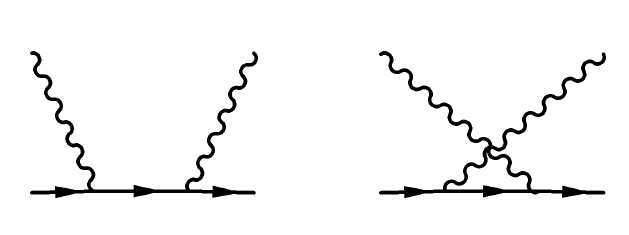}
    \includegraphics[height=0.15\textwidth]{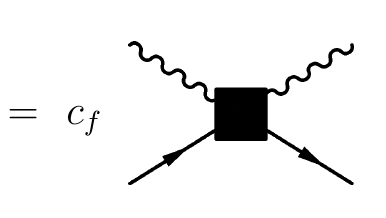}
    \vspace{5mm}
    \caption{Quark matching condition.   The right hand side denotes a sum
      over operators in Eq.~(\ref{eq:Tsymm}).
      \label{fig:quark}
    }
  \end{center}
\end{figure}

In order to match onto the quark operators, we compute forward
scattering amplitudes using  on-shell quarks with electromagnetic charge $e_f$ (in units of the proton charge), mass $m_f$,  and four-momentum $k$ as external states; cf. Fig.~\ref{fig:quark}.
The matrix element for $T_{\rm \sym}^{\mu\nu}$ between such states is 
\begin{align}
  \langle f|T_{\rm \sym}^{\mu\nu}|f \rangle
  =&-\sum_f\,\frac{2e_f^2}{q^2}\,
  \bar{u}_f(k)
  \left[ k^\mu\gamma^\nu+k^\nu\gamma^\mu -\frac{2k\cdot q}{q^2}
    \,\Big(q^\mu\gamma^\nu+q^\nu\gamma^\mu-g^{\mu\nu}\qslash\Big) \right]
u_f(k)
+\order\left(\frac1{q^4}\right)\,. 
\end{align}
After simplification this gives the results
\begin{equation}\label{eq:quarkmatch}
2c_{1\q}=c_{2\q}=-c_{3\q}=4e_{\q}^2 \,.
\end{equation}
The expression for $c_{1\q}$ agrees with equation 2.14 of Ref.~\cite{Collins:1978hi}.
The coefficients $c_{2\q}$ and $c_{3\q}$ were not considered in Ref.~\cite{Collins:1978hi}.

\subsection{Gluon operators}

\begin{figure}
\begin{center}
\includegraphics[height=0.15\textwidth]{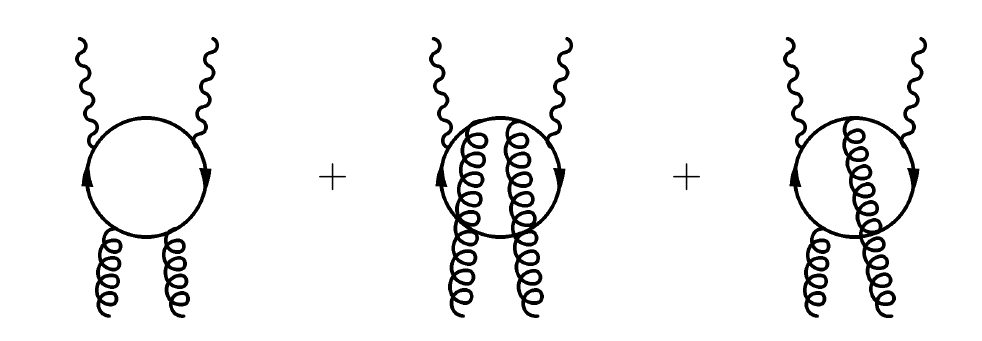}
\includegraphics[height=0.15\textwidth]{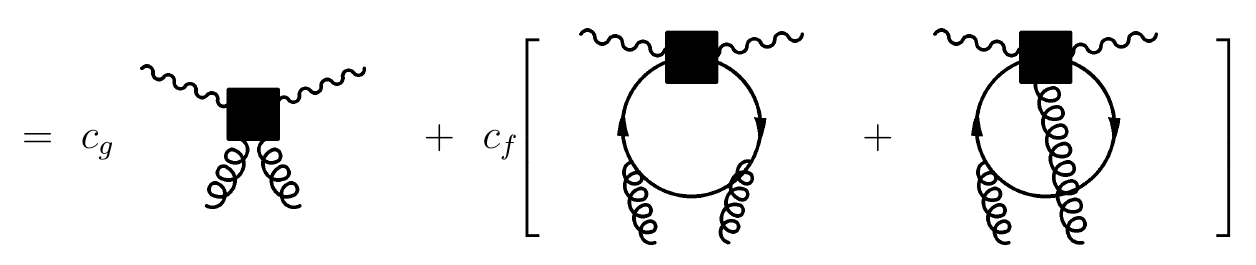}
\caption{\label{fig:gluonmatching}
  Matching condition onto gluon operators.  The three diagrams on the left hand side
  denote the contributions $T_a^{\mu\nu}$, $T_b^{\mu\nu}$ and $T_c^{\mu\nu}$ in Eq.~(\ref{gluon_tensor}). 
  The right hand side denotes a sum
  over operators in Eq.~(\ref{eq:Tsymm}).  Permutations of external states are not displayed.
  \label{fig:gluon}
}
\end{center}
\end{figure}

The calculation of $c_{ig}$ is simplified by using a background field analysis in
Fock-Schwinger gauge~\cite{Novikov:1983gd}.
We then have the following expressions for the gluon contributions to $T^{\mu\nu}$,
\begin{eqnarray}\label{gluon_tensor} 
  T_a^{\mu\nu}&=&-\dfrac{ig^2}{4}\mbox{Tr}\left[t^at^b\right]G^a_{\rho\alpha}(0)G^b_{\sigma\beta}(0)\sum_{\q}e^2_{\q}
  \int\dfrac{d^dl}{\left(2\pi\right)^d}\dfrac{\partial}{\partial k_\rho}\dfrac{\partial}{\partial k^\prime_\sigma}  \nonumber\\
  &&\mbox{Tr}\left[\gamma^\mu\dfrac{1}{\lslash-m_{\q}}\gamma^\alpha\dfrac{1}{\lslash-\kslash-m_{\q}}\gamma^\beta
    \dfrac{1}{\lslash-\kslash-\kslash^\prime-m_{\q}}\gamma^\nu\dfrac{1}{\lslash-\qslash-m_{\q}}\right]_{k=k^\prime=0} \,, \nonumber\\
  T_b^{\mu\nu}&=&-\dfrac{ig^2}{4}\mbox{Tr}\left[t^at^b\right]G^a_{\rho\alpha}(0)G^b_{\sigma\beta}(0)\sum_{\q}e^2_{\q}
  \int\dfrac{d^dl}{\left(2\pi\right)^d}\dfrac{\partial}{\partial k_\rho}\dfrac{\partial}{\partial k^\prime_\sigma}  \nonumber\\
  &&\mbox{Tr}\left[\gamma^\mu\dfrac{1}{\lslash-m_{\q}}\gamma^\nu\dfrac{1}{\lslash-\qslash+\kslash+\kslash^\prime-m_{\q}}
    \gamma^\alpha\dfrac{1}{\lslash-\qslash+\kslash^\prime-m_{\q}}\gamma^\beta\dfrac{1}{\lslash-\qslash-m_{\q}}\right]_{k=k^\prime=0} \,, \nonumber\\
  T_c^{\mu\nu}&=&-\dfrac{ig^2}{4}\mbox{Tr}\left[t^at^b\right]G^a_{\rho\alpha}(0)G^b_{\sigma\beta}(0)\sum_{\q}e^2_{\q}
  \int\dfrac{d^dl}{\left(2\pi\right)^d}\dfrac{\partial}{\partial k_\rho}\dfrac{\partial}{\partial k^\prime_\sigma} \nonumber\\
  &&\mbox{Tr}\left[\gamma^\mu\dfrac{1}{\lslash-m_{\q}}\gamma^\alpha\dfrac{1}{\lslash-\kslash-m_{\q}}
    \gamma^\nu\dfrac{1}{\lslash-\qslash+\kslash^\prime-m_{\q}}\gamma^\beta\dfrac{1}{\lslash-\qslash-m_{\q}}\right]_{k=k^\prime=0} \,,
\end{eqnarray}
where we separate $T^{\mu\nu}$ into
three pieces, $T_a^{\mu\nu}$, $T_b^{\mu\nu}$, and $T_c^{\mu\nu}$, as illustrated in Fig.~\ref{fig:gluon}. 
The contributions $T_a^{\mu\nu}$ and $T_b^{\mu\nu}$ are equal and IR divergent, while
$T_c^{\mu\nu}$ is IR finite. In the following we will work in
$d=4-2\epsilon$ dimensions  to regularize these IR divergences and
set the quark masses to zero.
The effective theory loop diagrams pictured in Fig.~\ref{fig:gluonmatching} then vanish. 
We note also that none of 
$T_a^{\mu\nu}$, $T_b^{\mu\nu}$ or $T_c^{\mu\nu}$ obeys current
conservation by itself, only their sum. 

Since $T_c^{\mu\nu}$ is IR finite we can set $d=4$ and get 
\begin{equation}
  T_c^{\mu\nu}=\sum_{\q}e^2_{\q}\,\dfrac{g^2}{16\pi^2}\dfrac1{q^2}\left[\dfrac13 O_g^{(0)}
    \left(-g^{\mu\nu}+\dfrac{q^\mu q^\nu}{q^2}\right)
    +2\dfrac{q^\mu q_\alpha}{q^2}O_g^{(2)\alpha\nu}
    +2\dfrac{q^\nu q_\alpha}{q^2}O_g^{(2)\alpha\mu}-4\dfrac{q_\alpha q_\beta}{q^2}O_{\rm \sym}^{(r)\mu\alpha\nu\beta} \right] \,,
\end{equation} 
where $q_\alpha q_\beta O_{\rm \sym}^{(r)\mu\alpha\nu\beta}$ is an operator structure whose
nucleon
matrix element vanishes, cf. Appendix~\ref{sec:appendix}.  
$T_a^{\mu\nu}$ is IR divergent and we calculate it  in $d=4-2\epsilon$ dimensions. We find 
\begin{eqnarray}
  T_a^{\mu\nu} = T_b^{\mu\nu} &=&\sum_{\q}e^2_{\q}\,\dfrac{g^2}{24\pi^2}\dfrac1{q^2}
  \left(\dfrac{4\pi\mu^2}{-q^2-i0}\right)^\epsilon \Gamma(2 + \epsilon)\mbox{B}(2 - \epsilon, 1 - \epsilon)\times\nonumber\\
  &\times&\bigg[\dfrac{q^\mu q^\nu q_\alpha q_\beta }{q^4}O_g^{(2)\alpha\beta}
    \left(-4-2\epsilon\right) +\dfrac{q_\alpha q_\beta g^{\mu\nu} }{q^2}O_g^{(2)\alpha\beta}
    \left( \dfrac{4 - 5\epsilon + 2\epsilon^2}{\epsilon}\right)\nonumber\\
    &&
    +\left(\dfrac{q_\alpha q^\mu}{q^2}O_g^{(2)\alpha\nu}+\dfrac{q_\alpha q^\nu}{q^2}O_g^{(2)\alpha\mu}\right)
    \left(\dfrac{-4+6\epsilon}{\epsilon}\right)+O_g^{(2)\mu\nu}
    \left(\dfrac{4-5\epsilon}{\epsilon\,(1+\epsilon)}\right)
    \bigg].
\end{eqnarray}
Expanding in $\epsilon$ and 
adding  $T_a^{\mu\nu}, T_b^{\mu\nu},$ and $T_c^{\mu\nu}$, 
we have the total gluon contribution
\begin{eqnarray}\label{total}
  T^{\mu\nu}_{\mbox{\scriptsize gluon}}
  &=&\sum_{\q}e^2_{\q}\,
  \dfrac{\alpha_s}{4\pi}
  \dfrac1{q^2}
  \Bigg[ \frac13 \left(-g^{\mu\nu}+\dfrac{q^\mu q^\nu}{q^2}\right)O_g^{(0)}
    -4 \dfrac{q_\alpha q_\beta}{q^2}O_{\rm \sym}^{(r)\mu\alpha\nu\beta}
    \nl
    &&\quad +
    \left( - \dfrac{8}{3 \epsilon} - \dfrac{14}{3} - \frac83 \log\dfrac{\mu^2}{\,-q^2\,}\right)
    \left(-g^{\mu\nu}+\dfrac{q^\mu q^\nu}{q^2}\right)\dfrac{q_\alpha q_\beta}{q^2}O_g^{(2)\alpha\beta}
    \nonumber\\
    &&\quad +
    \left( \dfrac{8}{3 \epsilon}+ 2 +  \frac83 \log\dfrac{\mu^2}{\,-q^2\,}\right)
    \left(-g^{\mu}_{\alpha}+\dfrac{q^\mu q_\alpha}{q^2}\right)\left(-g_{\beta}^{\nu}+\dfrac{q_\beta q^\nu}{q^2}\right)O_g^{(2)\alpha\beta}
    \Bigg] \,,
\end{eqnarray}  
from which we may read off the coefficients $c_{ig}$. 
Here we have introduced the renormalized strong coupling in the $\overline{\rm MS}$ scheme, 
\begin{align}
  g_{\rm bare}^2 \mu^{-2\epsilon}(4\pi)^{\epsilon}\Gamma(1+\epsilon)  = \alpha_s + \order(\alpha_s^2)
  \,.
\end{align}

As an alternative approach,
inspection of Eq.~(\ref{eq:Tsymm}) shows that the coefficients $c_{1}$, $c_{2}$ and $c_{3}$ may
be extracted from the quantity $v_\mu v_\nu T^{\mu\nu}$, for arbitrary timelike
unit vector $v^\mu$ ($v^2=1$):
\begin{align}\label{eq:Tvv}
  v_\mu v_\nu T_{\rm gluon}^{\mu\nu}
  &=
  c_{1g} {1\over q^2}  \left( -1 + {(v\cdot q)^2\over q^2} \right) O^{(0)}_g
  + 
  {q_\alpha q_\beta\over q^4} O^{(2)\alpha\beta}_g \left[ \left( -1 + {(v\cdot q)^2\over q^2} \right) c_{2g} +  { ( v\cdot q)^2\over q^2} c_{3g}
    \right]
  \nl
  &\quad 
  + c_{3g}\left( {v_\alpha v_\beta\over q^2} -2  { v\cdot q  q_\alpha v_\beta  \over q^4} \right) O^{(2)\alpha\beta}_g
  \,.
\end{align}
Following the notation of Ref.~\cite{Hill:2014yka}, we write%
\footnote{
  For our case, we need the ``ZZ"
  contribution from Ref.~\cite{Hill:2014yka}, with axial coefficient $c_A =0$ and vector coefficient $c_V=e_{\q}$. 
}
\begin{align}\label{eq:TfromI}
  v_\mu v_\nu T_k^{\mu\nu}
  &= -\dfrac{ig^2}{8} \bigg[ {1\over d(d-1)} O_g^{(0)} I_k^{(0)}(q)
    + {1\over d-2} O_g^{(2)\alpha\beta} I_{k\,\alpha\beta}^{(2)}(q) \bigg] \,. 
\end{align}
Neglecting quark masses, we have%
\footnote{
  A non-propagating typo appears in Eq.~(83) of Ref.~\cite{Hill:2014yka}:
  in $N_{a\,\mu\nu(ZZ)}^{(2)}$, in the $z_2 L_\mu L_\nu$ term, the opposite sign should appear in front of $x(2-x -\epsilon)$.  
}
\begin{align}
  I_a^{(0)}(\LL) &= 0 \,,
  \nl
  I_c^{(0)}(\LL) &= [c_\epsilon] \dfrac{\sum_f e_f^2}{ -\LL^2} \left[ 32 - 32{(v\cdot \LL)^2 \over \LL^2} \right] \,,
  \nl
  I_{a\,\mu\nu}^{(2)}(\LL) &= [c_\epsilon] {\sum_f e_f^2  \over (-\LL^2)^{1+\epsilon}} \bigg[
  v_\mu v_\nu \left( -{64\over 3\epsilon} + {16\over 3} \right)
  + {v\cdot \LL v_\mu \LL_\nu \over \LL^2}\left( {128\over 3\epsilon} + {64\over 3}\right)
  + {\LL_\mu \LL_\nu \over \LL^2} \left( -{64\over 3\epsilon} - 16 \right)
  \nl
  &\quad 
  + {\LL_\mu \LL_\nu (v\cdot \LL)^2 \over \LL^4} \left( {64\over 3}  \right) + \order(\epsilon)
  \bigg] \,,
  \nl
  I_{c\,\mu\nu}^{(2)}(\LL) &= [c_\epsilon] {\sum_f e_f^2  \over (-\LL^2)^{1+\epsilon}}
  \bigg[ -64 {v\cdot \LL v_\mu \LL_\nu \over \LL^2 } + \order(\epsilon) \bigg] \,, 
\end{align}
with $[c_\epsilon] \equiv i\Gamma(1+\epsilon)(4\pi)^{-2+\epsilon}$. 
Thus, either from Eq.~(\ref{total}) or from Eqs.~(\ref{eq:Tvv}) and (\ref{eq:TfromI}),
we read off the bare matching coefficients,
\begin{align} 
  c_{1g}^{\rm bare} &= {\alpha_s \over 4\pi} \sum_f e_f^2 \left(\frac13\right) \,,
  \nl
  c_{2g}^{\rm bare} &= {\alpha_s \over 4\pi} \sum_f e_f^2 \left( -q^2\over \mu^2 \right)^{-\epsilon}
  \left( -{8\over 3\epsilon} - {14\over 3} \right) \,,
  \nl
  c_{3g}^{\rm bare} &= {\alpha_s \over 4\pi} \sum_f e_f^2 \left( -q^2 \over \mu^2 \right)^{-\epsilon}
  \left( {8\over 3\epsilon} + 2 \right) \,. 
\end{align}
Renormalizing operators in the $\overline{\rm MS}$ scheme~\cite{Hill:2014yka}, we find the renormalized
coefficients, 
\begin{align}\label{eq:gluonmatch}
  c_{1g}(\mu) &= {\alpha_s \over 4\pi} \sum_f e_f^2 \left(\frac13\right) \,,
  \nl
  c_{2g}(\mu) &= {\alpha_s \over 4\pi} \sum_f e_f^2 
  \left(- {14\over 3}  + {8\over 3}\log{Q^2\over \mu^2} \right) \,,
  \nl
  c_{3g}(\mu) &= {\alpha_s \over 4\pi} \sum_f e_f^2 
  \left( 2 - {8\over 3}\log{Q^2\over \mu^2} \right) \,.
\end{align}

\section{Nucleon matrix elements} \label{Numerical}

Here we evaluate the relevant nucleon matrix elements.
In Secs.~\ref{sec:spin0me} and \ref{sec:spin2me}, we discuss
the spin-0 and spin-2 matrix elements for both proton and neutron.
In Sec.~\ref{sec:eval} we specialize to the proton case 
and provide explicit numerical results
for the leading $1/Q^2$ OPE evaluation of $W_1^p(0,Q^2)$. 

\subsection{Spin zero \label{sec:spin0me}}

Following Ref.~\cite{Hill:2014yxa} let us define%
\footnote{
  Compared to the quantity $f_{g,N}^{(0)}$ in Ref.~\cite{Hill:2014yxa},
  we have $\tilde{f}_{g,N}^{(0)} = [8\pi/(9\alpha_s)]f_{g,N}^{(0)}$.  The conventional $8/9$
  prefactor is designed to simplify the expression for the sum rule (\ref{eq:gluon}) when $n_f=3$.
}
\begin{align} \label{eq:scalar}
  \langle N(k) | O_{f}^{(0)}| N(k) \rangle &\equiv 2m_N^2 f_{f,N}^{(0)} \,,
  \quad 
      \langle N(k) | O_g^{(0)}(\mu) | N(k) \rangle 
      \equiv - 2 m_N^2 \tilde{f}_{g,N}^{(0)}(\mu) \,,
\end{align}
where we are using relativistic normalization $\bar{u}(k) u(k) = 2m_N$ for nucleon spinors.%
\footnote{
  Recall that the nonrelativistic normalization $\bar{u}(k) u(k) = m_N/E_{\bm{k}}$ was
  used in Ref.~\cite{Hill:2014yxa}. 
}
The gluon matrix element is determined by the mass sum rule, 
\begin{align}\label{eq:gluon}
  1
  = (1- \gamma_m) \sum_\q  f_{q,N}^{(0)} 
  - {\tilde{\beta}\over 2 } \tilde{f}_{g,N}^{(0)}  \,. 
\end{align} 
Here $\tilde{\beta}$ and $\gamma_m$ denote the QCD beta function and mass anomalous
dimension, whose leading expansions are 
\begin{align}\label{eq:beta}
  \tilde{\beta} &= {\beta\over g} = -\sum_{k=1} \beta_{k-1} \left(\alpha_s\over 4\pi\right)^k \,,
  \quad
  \beta_0 = 11-\frac23 n_f \,, 
  \nl
  \gamma_m &= -\sum_{k=1} \gamma_{k-1} \left(\alpha_s\over 4\pi\right)^k \,, \quad \gamma_0 = 8 \,, 
\end{align}
where $n_f$ is the number of active quark flavors.%
\footnote{Higher order terms are listed in Ref.~\cite{Hill:2014yxa}.  In numerical applications
  of Eq.~(\ref{eq:gluon}), we
  evaluate $\gamma_m$ and $\tilde{\beta}$ through $\order(\alpha_s^4)$. 
  }
For the quark matrix elements we use
\begin{align}\label{eq:sigmas}
  2m_N \Sigma_{\pi N} = {m_u+m_d\over 2} \langle N | (\bar{u} u + \bar{d} d) | N \rangle
  \,,
  \quad
2m_N \Sigma_{-} = (m_d - m_u) \langle N | (\bar{u} u - \bar{d} d) | N \rangle
\,. 
\end{align}
In terms of the quark mass ratio $R_{ud} \equiv {m_u / m_d}$ we have 
\begin{align}\label{eq:fufd}
f^{(0)}_{u,N} = {R_{ud} \over 1 + R_{ud}} \, {\Sigma_{\pi N} \over m_N} (1 + \xi) \, , \quad f^{(0)}_{d,N} = {1 \over 1 + R_{ud}} \, {\Sigma_{\pi N} \over m_N} (1 - \xi) \, , \quad \xi = {1 + R_{ud} \over 1 - R_{ud}}  \, { \Sigma_-  \over 2 \Sigma_{\pi N} } \,.
\end{align}
Numerical values for spin-zero matrix elements are summarized in Table~\ref{tab:spin0}.
As detailed in Ref.~\cite{Hill:2014yxa}, the matrix elements $f^{(0)}_{u,d,s}$ are identical
for $n_f=3$ and $n_f=4$, up to power corrections.  
\begin{table}[t]
\begin{center}
\begin{tabular}{l|rc}
quantity & value & reference \\
  \hline
  $\Sigma_{\pi N}$ & $44 (13) \,{\rm MeV}$ & \cite{Durr:2011mp,Durr:2015dna,Yang:2015uis,Abdel-Rehim:2016won,Bali:2016lvx,Hoferichter:2015dsa} \\
$\Sigma_-$ & $\pm 2(2)\,{\rm MeV}$ & \cite{Gasser:1982ap,Crivellin:2013ipa} \\
$R_{ud}$ &  $0.48(10)$ & \cite{Olive:2016xmw} \\
$m_N f^{(0)}_{s,N}$ & $40(20) \, {\rm MeV}$ & \cite{Junnarkar:2013ac} \\
$m_N f^{(0)}_{c,N}$  (lattice) & $80(30) \, {\rm MeV}$ &  \cite{Freeman:2012ry,Gong:2013vja,Abdel-Rehim:2016won} \\
$m_N f^{(0)}_{c,N}$  (pQCD) & $69(3) \, {\rm MeV}$ &  \cite{Hill:2014yxa} 
\end{tabular} 
\end{center}
\caption{\label{tab:spin0}
  Spin-zero operator matrix elements.  For $\Sigma_{\pi N}$, the value is taken from Ref.~\cite{Durr:2011mp}
  as in Ref.~\cite{Hill:2014yxa},
  and encompasses more recent results of
  Refs.~\cite{Durr:2015dna,Yang:2015uis,Abdel-Rehim:2016won,Bali:2016lvx,Hoferichter:2015dsa}.
  The upper (lower) sign in $\Sigma_-$ is for the proton (neutron). 
  The neutron form factors follow from approximate isospin symmetry expressed in (\ref{eq:PNiso}).
  For $f^{(0)}_{s,N}$  we assume a conservative $50 \%$ uncertainty compared to the estimate of $25 \%$
  in Ref.~\cite{Junnarkar:2013ac}.  For $f^{(0)}_{c,N}$ the value $80(30)\,{\rm MeV}$ encompasses the results of 
  Refs.~\cite{Freeman:2012ry,Gong:2013vja,Abdel-Rehim:2016won}.  
} 
\end{table}

\subsection{Spin two \label{sec:spin2me}} 

\begin{table}[t]
\begin{center}
  \begin{tabular}{c|c}
    quantity & value \\
    \hline
  $f_{u,p}^{(2)}(\mu=2\,{\rm GeV})$ & 0.346(6) \\
  $f_{d,p}^{(2)}(\mu=2\,{\rm GeV})$ & 0.192(5) \\
  $f_{s,p}^{(2)}(\mu=2\,{\rm GeV})$ & 0.034(2) \\ 
  $f_{c,p}^{(2)}(\mu=2\,{\rm GeV})$ & 0.0088(3) 
\end{tabular} 
\end{center}
\caption{\label{tab:momfraction}
  Spin-two proton matrix elements derived from MSTW analysis~\cite{Martin:2009iq}
  with $n_f=4$ at $\mu = 2\,{\rm GeV}$.
  The neutron form factors follow from approximate isospin symmetry expressed in Eq.~(\ref{eq:PNiso}).
} 
\end{table}

Let us define 
\begin{align}\label{eq:tensor}
\langle N(k) | O_{\q}^{(2)\mu\nu}(\mu) | N(k) \rangle 
&\equiv 2 \left(k^\mu k^\nu - {g^{\mu\nu} \over 4} m_N^2 \right) f_{\q,N}^{(2)}(\mu) \,,
\nl
\langle N(k) | O_g^{(2)\mu\nu}(\mu) | N(k) \rangle 
&\equiv 2 \left(k^\mu k^\nu - {g^{\mu\nu} \over 4} m_N^2 \right) f_{g,N}^{(2)}(\mu) \,.
\end{align} 
The gluon matrix element is determined by the momentum sum rule, 
\begin{align}\label{eq:s2sum}
\sum_{\q} f_{\q,N}^{(2)}(\mu) + f_{g,N}^{(2)}(\mu) = 1 \,.
\end{align}
Up to corrections proportional to $m_u-m_d$ and $\alpha$, the proton and neutron matrix elements are related as, 
\be\label{eq:PNiso}
\langle p | O^{(2)}_{u} | p \rangle = \langle n | O^{(2)}_{d} | n \rangle \, , \quad \langle p | O^{(2)}_{d} | p \rangle = \langle n | O^{(2)}_{u} | n \rangle \, , \quad \langle p | O^{(2)}_{s} | p \rangle = \langle n | O^{(2)}_{s} | n \rangle \, .
\ee
For the numerical evaluation, we use the inputs of Table~\ref{tab:momfraction} for $n_f=4$ at $\mu=2\,{\rm GeV}$,
with gluon matrix element given by Eq.~(\ref{eq:s2sum}).
For different $\mu$ in the limited range
considered, we employ the leading log renormalization (cf. Table 5 of Ref.~\cite{Hill:2014yxa}).

\subsection{Large $\bm{Q^2}$ behavior of $\bm{ W_1(0,Q^2)}$: numerical results \label{sec:eval}}

Let us restrict attention to the subtraction function for the proton, $W_1^p(0,Q^2)$. 
For notational simplicity we suppress the superscript, $W_1^p \to W_1$, in the following.  
In terms of the matrix elements of the previous section, we have the leading power result
\begin{align}
  W_1(0,Q^2) &= {2m_p^2\over Q^2} \bigg\{
  -\sum_f c_{1f} f_f^{(0)} + c_{1g}
  \tilde{f}_{g}^{(0)} 
  +\frac14 \bigg[ \sum_f \left( c_{2f} - c_{3f} \right) f^{(2)}_f
    + \left( c_{2g} - c_{3g} \right) f^{(2)}_g \bigg]
    \bigg\} \,, 
\end{align}
where the first nonvanishing order for coefficients of each operator are given by
Eqs.~(\ref{eq:quarkmatch}) and (\ref{eq:gluonmatch}). 
Let us consider separately the spin-0 and spin-2 contributions.

\subsubsection{Spin-0: numerical evaluation for proton}

\begin{figure}[t]
\centering
\includegraphics[height=0.4\textwidth]{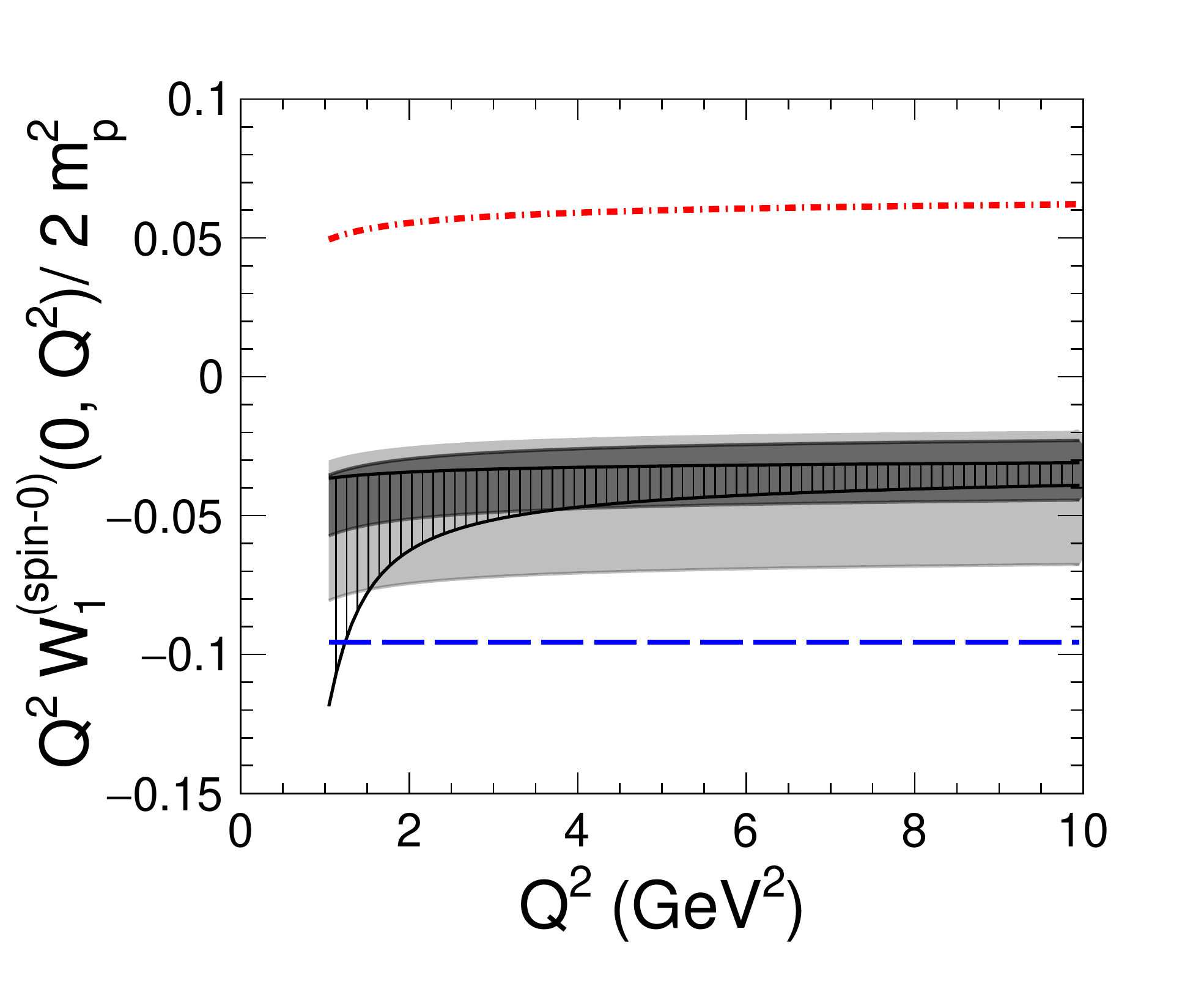}
\includegraphics[height=0.4\textwidth]{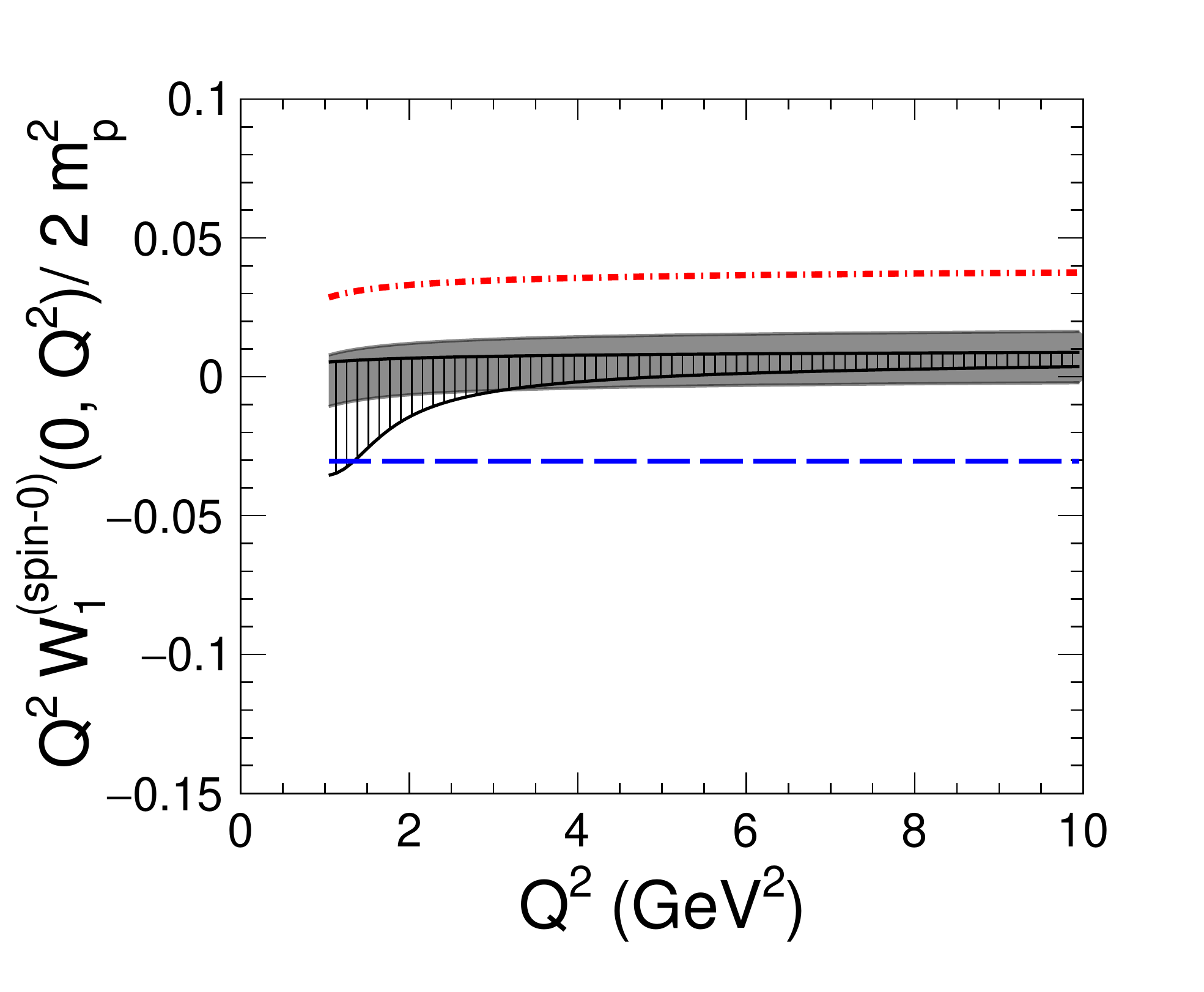}
\caption{Spin-0 contribution to $Q^2\,W_1(0,Q^2)/2m_p^2$ as a function of $Q^2$.
  The final results are given with $n_f=4$ in the left-hand plot.  For comparison, results with
  $n_f=3$ are in the right-hand plot.  
  The black band with vertical stripes 
  displays the result with perturbative uncertainty estimated by renormalization scale variation
  $Q/2 < \mu < 2Q$.  The solid bands represent hadronic uncertainty from
  local matrix elements in Table~\ref{tab:spin0}. 
  In the $n_f=4$ evaluation, dark and light solid bands are obtained using 
  $f_{c}^{(0)}(\rm pQCD)$ and $f_{c}^{(0)}(\rm lattice)$, respectively. 
  The bottom dashed blue line and
  top dash-dotted red line give the separate quark and gluon contributions to the central
  value.  
  \label{fig:spin0}}
\end{figure}

Substituting the Wilson coefficients from Eqs.~(\ref{eq:quarkmatch}) and (\ref{eq:gluonmatch}), 
the spin-0 contribution is
\begin{align} \label{eq:spin0}
  {Q^2\over 2m_p^2} W_1^{({\rm spin}-0)}(0,Q^2) &=
  - 2 \sum_f e_f^2 f_f^{(0)}
  + \left(\sum_f e_f^2\right) {\alpha_s\over 12\pi} \tilde{f}_g^{(0)}(\mu)
  \,. 
\end{align}
This contribution is displayed in Fig.~\ref{fig:spin0} using
$n_f=4$ and $n_f=3$.  
Having neglected quark masses, our $n_f=4$ result is formally valid in
the $m_c^2 \ll Q^2$ regime.
Alternatively, an evaluation using $n_f=3$ is formally valid in the regime
$\Lambda_{\rm QCD}^2 \ll Q^2 \ll m_c^2$.
We show below that $W_1(0,Q^2)$ is numerically dominated by the spin-2
contribution, and do not pursue a more elaborate analysis of charm quark mass
dependence in the spin-0 contribution.   

We take $n_f=4$ in our final result.  Fig.~\ref{fig:spin0} compares results using different
values of the charm scalar matrix element, $f_c^{(0)}$.   As default, we use
the perturbative QCD value obtained from expanding in $\Lambda_{\rm QCD}/m_c$~\cite{Hill:2014yxa}.
For comparison,
Fig.~\ref{fig:spin0} displays the range of $f_c^{(0)}$ from 
currently available $n_f=4$ lattice QCD evaluations~\cite{Freeman:2012ry,Gong:2013vja,Abdel-Rehim:2016won}.
The gluon matrix element is given by Eq.~(\ref{eq:gluon}),
with $\tilde{\beta}$ and $\gamma_m$ evaluated through $\order(\alpha_s^4)$.  
We choose default renormalization scale $\mu=Q$, and 
perturbative scale uncertainty is estimated by varying $Q/2 < \mu < 2 Q$.
We remark that a partial cancellation occurs between quark and gluon matrix elements,
as illustrated in the figure.  

The spin-0 part disagrees with Collins's calculation~\cite{Collins:1978hi}. 
Equation 2.18 of Ref.~\cite{Collins:1978hi} is 
\begin{equation}
  W_1^{({\rm spin}-0)}(0,Q^2)^{\rm Collins}
  =\dfrac1{Q^2}\left\{\dfrac{4KM^2}{a}-\sum_i\left(\dfrac{2K}{a}+2\kappa_i^2\right)\langle p|m_{iB}\bar\psi_i\psi_i|p\rangle \right\},
\end{equation}
where $M=m_p$ is the proton mass,
$K/a = n_f/(3\beta_0)$ with $\beta_0$ given by Eq.~(\ref{eq:beta}), 
$\kappa_i$ is the quark charge (which we have denoted $e_i$),
and $m_{Bi}$ is the bare quark mass.
In our notation,
\begin{equation}\label{eq:Coll}
    W_1^{({\rm spin}-0)}(0,Q^2)^{\rm Collins}=\dfrac{2m_p^2}{Q^2}\left\{\dfrac{2n_f}{3\beta_0}-
  \sum_i \left(\dfrac{2n_f}{3\beta_0} + 2 e_i^2 \right) f_i^{(0)} \right\}  \,. 
\end{equation}  
Comparing with our result for the spin-0 part (\ref{eq:spin0}), and using the
relation (\ref{eq:gluon}) with $\gamma_m=0$ and $\tilde{\beta}=-\alpha_s\beta_0/(4\pi)$,
we see that the replacement $n_f\to {\sum_{\q=u,d,s} e_{\q}^2}\,$
must be made in Collins's expression (\ref{eq:Coll}) to obtain the correct result. 

\subsubsection{Spin-2: numerical evaluation for proton}

\begin{figure}[htb]
\centering
\includegraphics[height=0.5\textwidth]{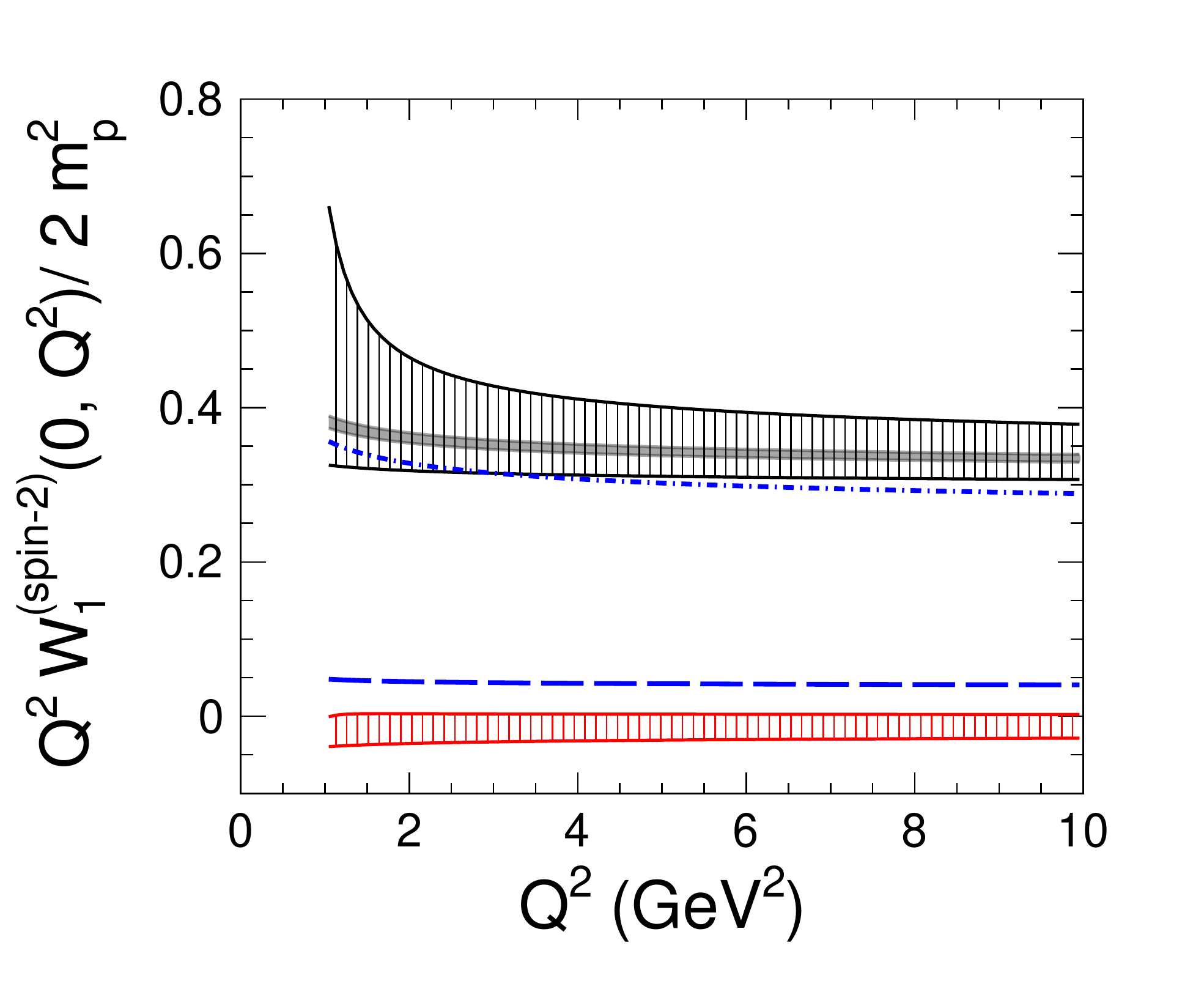}
\caption{Spin-2 contribution to $Q^2\,W_1(0,Q^2)/2m_p^2$ as a function of $Q^2$.
  The dark gray band represents hadronic uncertainty. 
  The vertically hatched band represents perturbative uncertainty, estimated by
  varying renormalization scale $Q/2 < \mu < 2Q$. 
  Upper dash-dotted and lower dashed blue lines represent the contributions of
  up-quark and down-quark to the central value, and the lower red hatched band shows
  the gluon contribution and uncertainty. 
  \label{fig:spin2}}
\end{figure}

Substituting the Wilson coefficients from Eqs.~(\ref{eq:quarkmatch}) and (\ref{eq:gluonmatch}), 
the spin-2 contribution is  
\begin{align}
\label{eq:spin2}
  {Q^2\over 2m_p^2} W_1^{({\rm spin}-2)}(0,Q^2) &=
   2 \sum_f e_f^2 f_f^{(2)}
   + \left(\sum_f e_f^2\right)  {\alpha_s\over 4\pi} \left( -\frac53 + \frac43 \log{Q^2\over\mu^2} \right) f_g^{(2)}(\mu) \,,
\end{align}
where we retain the first non-vanishing contribution to each operator matching coefficient. 
This result is displayed in Fig.~\ref{fig:spin0}.  For definiteness, we employ the $n_f=4$ results from
Table~\ref{tab:momfraction}, although the contributions from strange and charm quark are negligible.
The result is dominated by the up quark, with small contributions from down quark and gluon;
the latter contributes a sizable fraction of the uncertainty as indicated in the figure.  

\subsubsection{Total \label{sec:W1tot}}

\begin{figure}[htb]
\centering
\includegraphics[height=0.5\textwidth]{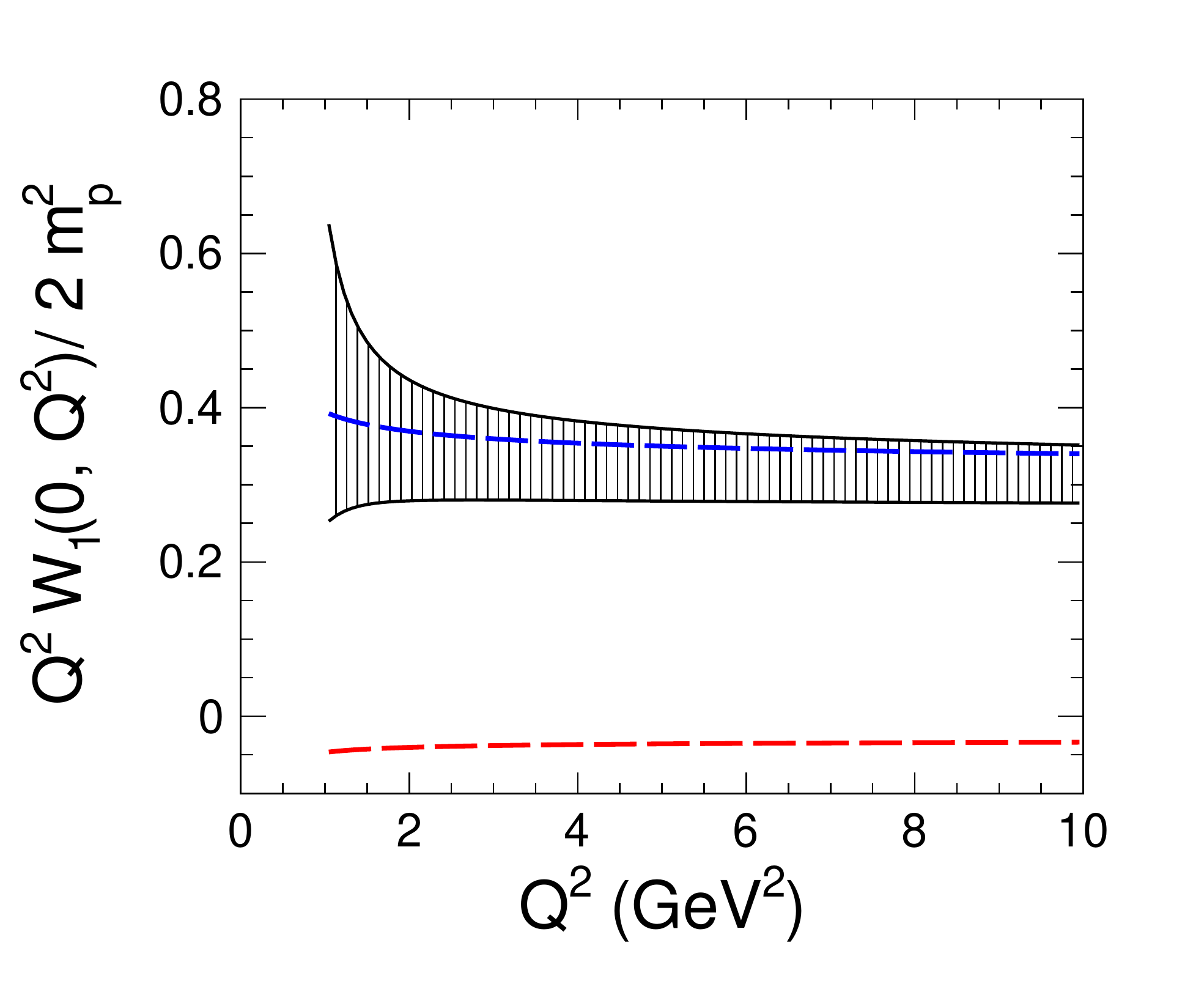}
\caption{Total $Q^2\,W_1(0,Q^2)/2m_p^2$ as a function of $Q^2$.
  The dashed red and blue lines denote central values for the spin-0 and spin-2 contributions. 
  The vertical hatched black band is the total contribution, with perturbative and
  hadronic errors from each of the spin-0 and spin-2 contributions added in quadrature. 
  \label{fig:tot}}
\end{figure}

The complete OPE evaluation of $W_1(0,Q^2)$ for the proton
is given by the sum of spin-0 and spin-2 contributions, and
is displayed in Fig.~\ref{fig:tot}.
The result is dominated by the spin-2 contribution. 

\section{Two photon exchange in muonic hydrogen Lamb shift} \label{Applications}

As detailed in Ref.~\cite{Hill:2011wy}, the piece of the TPE
contribution to the muonic hydrogen Lamb shift
that is not fully determined by electron
scattering data involves the subtraction function in the dispersion relation for the forward Compton
amplitude of the proton, i.e. $W_1(0,Q^2)$.  Having constrained this function at
both high- and low-$Q^2$, let us consider the implications for 
muonic hydrogen.  Here we construct an interpolation of this function with the
correct coefficients of $(Q^2)^0$ and $(Q^2)^1$ at low $Q^2$, the correct
coefficient of $1/Q^2$ at high $Q^2$, and higher order terms
parameterized by $\Lambda_L$ and $\Lambda_H$ introduced below.  

Recall that at low-$Q^2$, we have the expansion,~\cite{Hill:2011wy}
\begin{multline}\label{eq:lowQ2}
  W_1(0,Q^2)_L = 2a_p (2+a_p)  \\
  + 
  {Q^2\over m_p^2}
  \bigg\{ {2m_p^3 \bar{\beta} \over \alpha} -  a_p
  - \frac23\bigg[ (1+a_p)^2 m_p^2(r_M^p)^2 - m_p^2(r_E^p)^2 \bigg] \bigg\} 
  \bigg[ 1 + \Delta_L(Q^2) \bigg] \,,
\end{multline} 
where the anomalous magnetic moment of the proton is $a_p = 1.793$, 
the magnetic polarizability of the proton%
\footnote{
See e.g. Ref.~\cite{Hill:2012rh} for the definition of $\bar{\beta}$, which is there denoted $\beta_M$. 
}
is $\bar{\beta}= 2.5(4)\times 10^{-4}\,{\rm fm}^3$~\cite{Olive:2016xmw},
and the proton magnetic and electric radii are 
$r_M = 0.776(34)(17)\,{\rm fm}$~\cite{Lee:2015jqa,Olive:2016xmw}, $r_E=0.8751(61)\,{\rm fm}$~\cite{Mohr:2015ccw,Olive:2016xmw}.%
\footnote{After accounting for other uncertainties,
  the results of the interpolation for $\delta E(2S)^{W_1(0,Q^2)}$ are not significantly
  altered by using the muonic hydrogen value, $r_E=0.84087(26)(29)\,{\rm fm}$~\cite{Antognini:1900ns,Olive:2016xmw}.  
  }
Here $\Delta_L(Q^2)$ accounts for terms of order $Q^4$ and higher in the Taylor expansion.
To investigate sensitivity to these terms we consider
\begin{align}\label{eq:curve}
 \Delta_L(Q^2) = \pm {Q^2\over \Lambda_L^2} \,,
\end{align}
with hadronic scale $\Lambda_L \approx 500\,{\rm MeV}$. 

\begin{figure}[htb]
  \centering
  \includegraphics[height=0.5\textwidth]{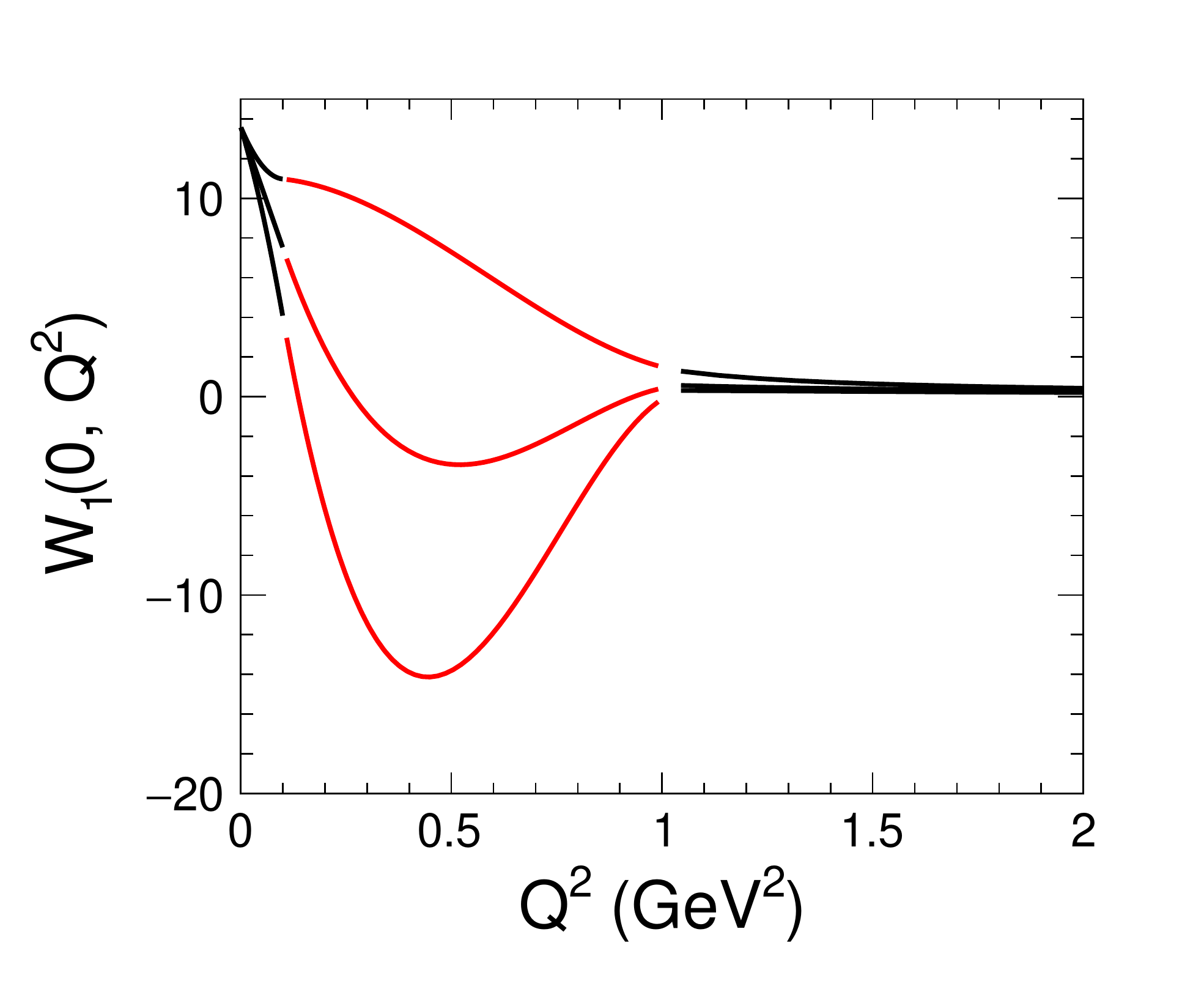}
  \caption{$W_1(0,Q^2)$ as a function of $Q^2$.
    Solid black curves on the left and right are from Eqs.~(\ref{eq:lowQ2}) and (\ref{eq:highQ2}),
    as explained in the text.  Intermediate red curves are interpolations from the central curves
    and envelopes.
    \label{fig:W1_interp}}
\end{figure}

For the high-$Q^2$ behavior we have 
\begin{align}\label{eq:highQ2}
 W_1(0,Q^2)_H = {2m_p^2\over Q^2} A(Q^2) \bigg[ 1 + \Delta_H(Q^2) \bigg] \,, 
\end{align}
where $A(Q^2)$ is the leading power result displayed in Fig.~\ref{fig:tot},
and $\Delta(Q^2)$ accounts for power corrections (including charm mass corrections).  
To investigate the impact of power-suppressed terms we take central
value $\Delta_H=0$, and upper and lower envelopes 
\begin{align}\label{eq:power}
  \Delta_H(Q^2) =  \pm {\Lambda_H^2\over Q^2} \,,
\end{align}
with hadronic scale $\Lambda_H \approx 500\,{\rm MeV}$.  
Fig.~\ref{fig:W1_interp} displays
a cubic spline interpolation of $W_1(0,Q^2)$
taking the low-$Q^2$ and high-$Q^2$ inputs from NRQED and OPE.
We show results for central values, and for the interpolations based on
upper and lower envelopes for the low-$Q^2$ ($Q^2<Q_L^2$) and high-$Q^2$ ($Q^2>Q_H^2$)
constraints.  For our default illustration we take $Q_L^2=0.1\,{\rm GeV}^2$ and
$Q_H^2=1\,{\rm GeV}^2$.  
At low-$Q^2$ these envelopes account for uncertainties in low-energy constants
appearing in Eq.~(\ref{eq:lowQ2}), and for curvature and higher order corrections
according to Eq.~(\ref{eq:curve}).%
\footnote{
  Choosing a lower cutoff $Q_L^2=0.05\,{\rm GeV}^2$ would
  give $[-0.043\,{\rm meV}, -0.023\,{\rm meV}]$ in place of Eq.~(\ref{eq:deltaEtot}). 
  }
At high-$Q^2$ these envelopes account for perturbative
and hadronic matrix element uncertainties in $A(Q^2)$ from Fig.~\ref{fig:tot},
and for power corrections according to Eq.~(\ref{eq:power}).%
\footnote{
  Choosing a higher cutoff $Q_H^2=2\,{\rm GeV}^2$ would give $[-0.052\,{\rm meV},\, -0.020\,{\rm meV}]$
  in place of Eq.~(\ref{eq:deltaEtot}).  
  }
The intermediate region is outside of the domain of validity of either
NRQED
at low-$Q^2$, or OPE at high-$Q^2$. 

\begin{figure}[t]
  \begin{center}
    \includegraphics[height=0.5\textwidth]{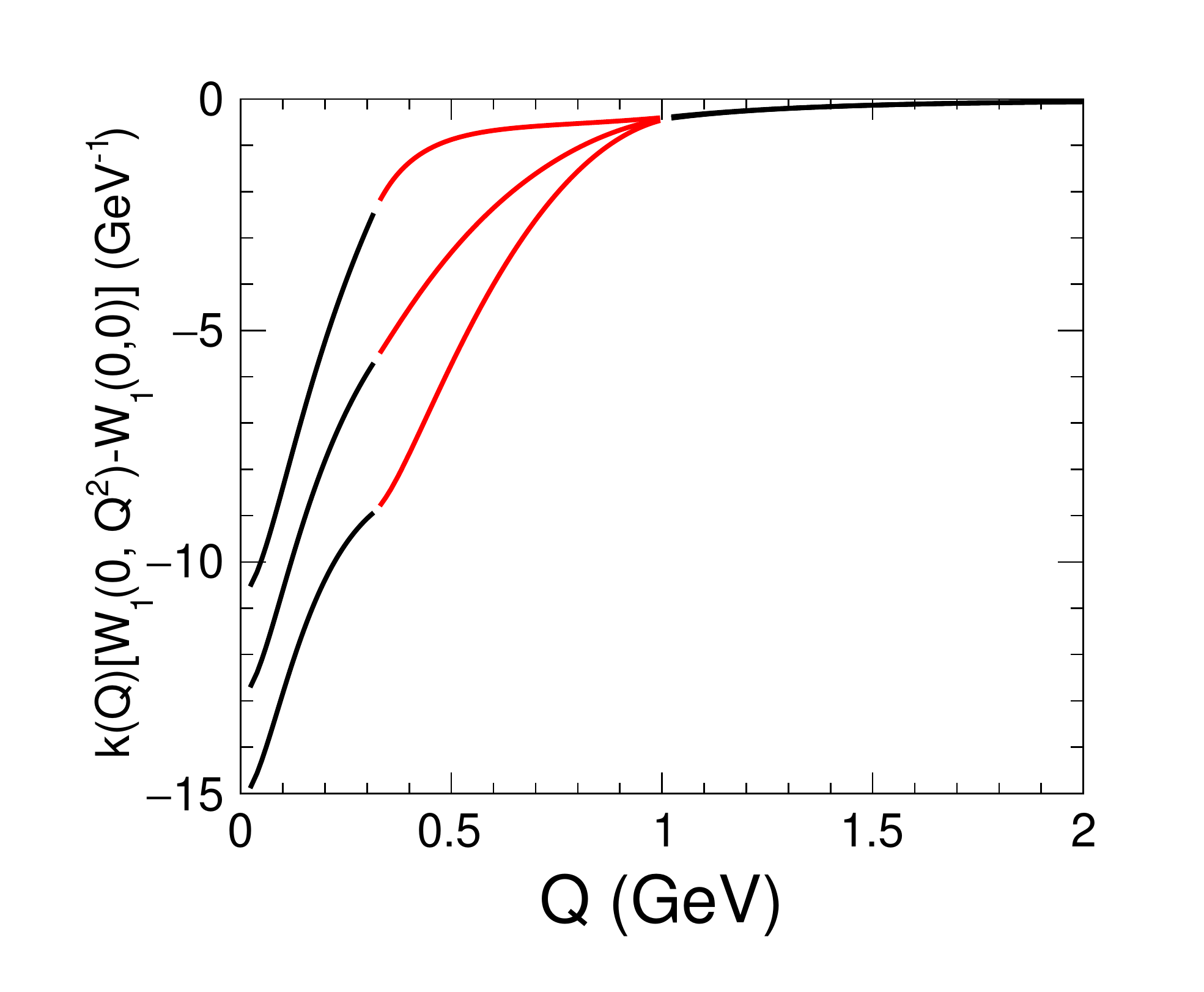}
    \caption{$W_1(0,Q^2)$, after subtracting $Q^2=0$ limit and scaling by the
      kinematic factor $k(Q)$ in Eq.~(\ref{eq:kq}), as a function of $Q$.  The contribution
      to the muonic hydrogen Lamb shift is proportional to the area under the curve,
      cf. Eq.~(\ref{eq:shift}).
      \label{fig:W1_moment}}
  \end{center}
\end{figure}

The relevant TPE contribution to muonic hydrogen may be written
\begin{align}\label{eq:shift}
  \delta E({\rm 2S})^{W_1(0,Q^2)} = { \alpha^2 \over m_\mu m_p} |\psi_{\rm 2S}(0)|^2
  \int_0^\infty dQ \, k(Q) \big[ W_1(0,Q^2) - \lim_{Q^2 \to 0}W_1(0,Q^2) \big] \,,
\end{align}
where $|\psi_{2S}(0)|^2 = m_r^3\alpha^3/(8 \pi)$ for the 2S atomic state, 
and $k(Q)$ is the kinematic function 
\begin{align}\label{eq:kq}
  k(Q) = {4\over \pi} {m_\mu^2\over Q} \int_{-1}^1 dx \, { \sqrt{1-x^2} (1+2x^2) \over Q^2 + 4m_\mu^2 x^2} \,.
\end{align}
After subtracting the $Q^2=0$ limit,
and rescaling with $k(Q)$, the interpolated function
is displayed in the LHS of Fig.~\ref{fig:W1_moment}.
The energy shift in Eq.~(\ref{eq:shift}) is proportional to the area under the curve in Fig.~\ref{fig:W1_moment}.
For the central curve we obtain $\delta E({\rm 2S})^{W_1(0,Q^2)} = -0.034\,{\rm meV}$, and for the
interpolated envelopes displayed in the figure, we find the interval 
\begin{align} \label{eq:deltaEtot}
  \delta E({\rm 2S})^{W_1(0,Q^2)}\big|_{\rm Fig.~\ref{fig:W1_moment}} \in [ -0.046\,{\rm meV} , \,  -0.021\,{\rm meV}]  \,.
\end{align}

We present in Table~\ref{tab:conts}
the total structure-dependent TPE contribution
to the $2P-2S$ energy splitting in muonic hydrogen.
Following the terminology of Ref.~\cite{Hill:2011wy} (cf. Table~I of this reference), 
we write the total contribution 
as a sum of proton-pole, $W_1(0,Q^2)$, and continuum, contributions.
Since our focus is on the impact of $W_1(0,Q^2)$, for simplicity we retain a simple dipole ansatz for the
proton pole contributions.%
\footnote{As in Ref.~\cite{Hill:2011wy}, we take $G_E(q^2)=G_M(q^2)/G_M(0)=[1-q^2/\Lambda^2]^{-2}$, $\Lambda^2=0.71\,{\rm GeV}^2$.
  A more systematic evaluation of the proton pole contribution and uncertainty is considered in Ref.~\cite{ff}.
}
The continuum contribution is from Ref.~\cite{Carlson:2011zd}. 

\begin{figure}[htb]
  \centering
  \includegraphics[height=0.5\textwidth]{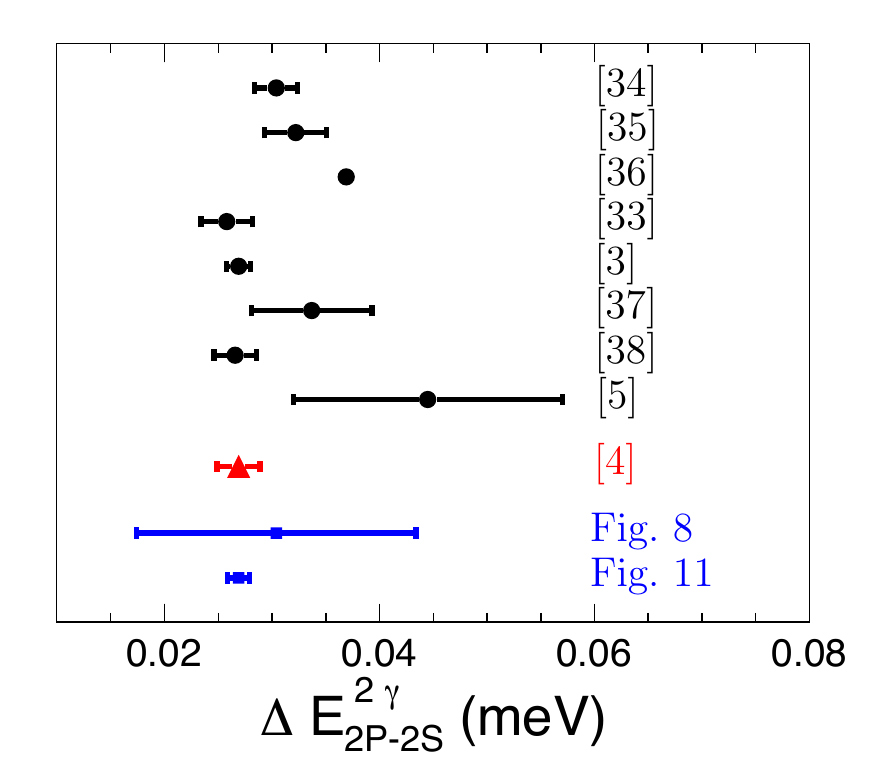}
  \caption{Two-photon contribution to the Lamb shift in muonic hydrogen,
    adjusted to common proton elastic form factors (see text). 
    Black circles denote previous work, blue square denotes the present work.
    The red triangle is the summary of Ref.~\cite{Antognini:2013jkc} used
    in the CREMA muonic hydrogen extraction of $r_E^p$.  
    \label{fig:DeltaE}}
\end{figure}

It is often conventional to subtract off a contribution to TPE corresponding to the
``Sticking In Form Factors'' (SIFF) ansatz~\cite{Hill:2011wy}.
With the standard dipole form factors, this contribution is%
\footnote{
  In the literature, this may be conventionally
  referred to as an ``elastic'', ``Born'', or ``proton'' contribution.
}
\begin{align}\label{eq:SIFFpart}
\delta E(2P-2S)^{ {\rm two}-\gamma}\big|_{\rm SIFF} = 0.0184\,{\rm meV} \,. 
\end{align}
In order to make a meaningful comparison with previous work, we add this contribution
back to results where it has been subtracted.%
\footnote{
  Thus we add $\delta E(2P-2S)^{ {\rm two}-\gamma}\big|_{\rm SIFF}$ from Eq.~(\ref{eq:SIFFpart})
  to the ``polarizability'' of Refs.~\cite{Pachucki:1999zza,Martynenko:2005rc,Nevado:2007dd,Alarcon:2013cba,Gorchtein:2013yga},
  (for Ref.~\cite{Gorchtein:2013yga} we show the adjusted value quoted in Ref.~\cite{Alarcon:2013cba}), 
  and to the sum or ``subtraction'' and ``inelastic'' terms of Ref.~\cite{Carlson:2011zd}. 
  In Refs.~\cite{Birse:2012eb,Antognini:2013jkc} we take $\Delta E_{\rm el} + \Delta E_{\rm np}
  =0.0247\,{\rm meV} \to 0.0184\,{\rm meV}$.  
  In Ref.~\cite{Peset:2014jxa}, we take $\Delta E_{\rm Born} = 0.0083\,{\rm meV} \to
  0.0184\,{\rm meV}$.
}
The resulting total structure-dependent
TPE correction from a range of approaches is displayed in Fig.~\ref{fig:DeltaE}. 

\begin{table}[htb]
\caption{\label{tab:conts} 
  Two-photon proton structure corrections to the $2P-2S$ Lamb shift in muonic hydrogen, in meV. }
\begin{center}
\begin{tabular}{c|l}
  Contribution & value (meV) \\
  \hline\hline
$\delta E^{\rm proton}_{\mu H}$& \hspace{-0.5mm} $-0.016$ \\
\hline
$\delta E^{W_1(0,Q^2)}_{\mu H}$ & \quad $0.034(13)$ \\
\hline
$\delta E^{\rm continuum}_{\mu H}$ & \quad $0.0127(5)$ \\
\hline\hline 
$\delta E^{{\rm two}-\gamma}$ & \quad $0.030(13)$ 
\end{tabular} 
\end{center}
\end{table}

\section{Conclusions}  \label{Conclusions}

The forward Compton amplitude is an important ingredient for
problems in particle, nuclear and atomic physics involving nucleon structure.
The subtraction function $W_1(0,Q^2)$ is a key
source of uncertainty.
Neither the spin-0 contribution, nor the spin-2 contribution  
to the leading OPE expression for this function
has been correctly included previously. 
We computed this OPE expression and employed it to constrain the 
TPE contribution to the muonic-hydrogen Lamb shift. 

The evaluation of $W_1(0,Q^2)$ by OPE was considered long ago by
Collins~\cite{Collins:1978hi}.  This reference considered the
electromagnetic corrections to the nucleon mass, where only the spin-0
contribution is relevant. As described in Sec.~\ref{sec:eval}, an incorrect
expression in Ref.~\cite{Collins:1978hi} impacts the isoscalar spin-0
contribution.  The spin-2 contribution was not considered in
Ref.~\cite{Collins:1978hi}, and is found numerically to dominate the
total OPE expression for $W_1^p(0,Q^2)$. For the application to the
proton-neutron mass difference, both the spin-2 and the isoscalar
spin-0 parts drop out.%
\footnote{
These observations imply that the electromagnetic contribution to the neutron-proton
mass difference cannot be used to directly constrain
the TPE contribution to muonic hydrogen spectroscopy.
}
However, for the application to muonic
hydrogen spectroscopy these ingredients are critical.

Using a parameterization of higher order corrections, Eqs.~(\ref{eq:curve}) and (\ref{eq:power}),
an interpolation of $W_1(0,Q^2)$ with the correct low-$Q^2$ and high-$Q^2$ behavior is displayed
in Fig.~\ref{fig:W1_interp}.  The corresponding TPE contribution and uncertainty
for the Lamb shift in muonic hydrogen is given by the area under the curve in
Fig.~\ref{fig:W1_moment}, as displayed in Fig.~\ref{fig:DeltaE}. 
This uncertainty remains the dominant one in the muonic hydrogen
Lamb shift, and for derived observables such as the proton charge radius
and Rydberg constant using muonic hydrogen inputs. 
Further constraints arising from chiral lagrangian analysis
in the $m_\pi^2 \lesssim Q^2 \ll m_p^2$ regime are discussed in Appendix~\ref{sec:chiral}. 
Obtaining an error bar of the size assumed in the CREMA analysis~\cite{Antognini:1900ns} (the point corresponding to Ref.~\cite{Antognini:2013jkc} in Fig.~\ref{fig:DeltaE})
requires some level of model dependence. 
Future measurements with light muonic atoms and questions
surrounding the proton radius puzzle may motivate direct computations
using lattice QCD. 

\vskip 0.2in
\noindent
{\bf Acknowledgments}
\vskip 0.1in
\noindent
We  thank J.~Arrington, A.~Blechman,  G.~Lee, A.~Majumder, A.~Walker-Loud, I.~Yavin and Z.~Ye for useful discussions.
G.P. thanks Perimeter Institute for Theoretical Physics, where part of this work was done,
for its hospitality and support. This work was supported by NIST Precision Measurement
Grants Program (R.J.H., G.P.), DOE grant DE-FG02-13ER41958 (R.J.H.) and DOE grant DE-SC0007983 (G.P.). 
Research at Perimeter Institute is supported by the Government of Canada through the Department of Innovation,
Science and Economic Development and by the Province of Ontario through the Ministry of Research and Innovation.
TRIUMF receives federal funding via a contribution agreement with the National Research Council of Canada.
Fermilab is operated by Fermi Research Alliance, LLC under Contract No. DE-AC02-07CH11359 with the United States
Department of Energy. 

\begin{appendix}
  
  \section{Details on gluon operator structures \label{sec:appendix} }

In the calculation we encounter the structure $q_\alpha q_\beta G^{A\mu\alpha}G^{A\nu\beta}$. $G^{A\mu\alpha}G^{A\nu\beta}$ is the color-singlet product of a chromo-electric ($E_i^A$)  and a chromo-magnetic ($B_i^A$) field. As such, it has $7\times6/2=21$ linearly independent components. Since $q_\alpha q_\beta G^{A\mu\alpha}G^{A\nu\beta}$ depends only on the symmetric part of  $G^{A\mu\alpha}G^{A\nu\beta}$ under $\alpha\leftrightarrow\beta$, it can be written in terms of the four-index tensor $O_{\rm \sym}^{\mu\alpha\nu\beta}\equiv\frac12\left(G^{A\mu\alpha}G^{A\nu\beta}+G^{A\mu\beta}G^{A\nu\alpha}\right)$
as $q_\alpha q_\beta O_{\rm \sym}^{\mu\alpha\nu\beta}$. The tensor $O_{\rm \sym}^{\mu\alpha\nu\beta}$ has the
index symmetries $O_{\rm \sym}^{\mu\alpha\nu\beta}=O_{\rm \sym}^{\mu\beta\nu\alpha}=O_{\rm \sym}^{\nu\beta\mu\alpha}$
and 20 linearly independent components. Nine of these can be identified with the components of $O_g^{(2)\alpha\beta}$ and one with $O_g^{(0)}$. 

Following \cite{Hill:2014yka}, define 
\begin{eqnarray}
 O^{(0)\mu\alpha\nu\beta}&=&\frac{1}{d(d-1)}O_g^{(0)}\left(g^{\mu\nu}g^{\alpha\beta}-g^{\mu\beta}g^{\nu\alpha}\right) \,, \nonumber\\
 O^{(2)\mu\alpha\nu\beta}&=&\frac{1}{d-2}\left(g^{\mu\beta}O_g^{(2)\nu\alpha}+g^{\nu\alpha}O_g^{(2)\mu\beta}
  -g^{\mu\nu}O_g^{(2)\alpha\beta}-g^{\alpha\beta}O_g^{(2)\mu\nu}\right), 
\end{eqnarray} 
and  $O_{\rm \sym}^{(0)\mu\alpha\nu\beta}=\frac12\left(O^{(0)\mu\alpha\nu\beta}+O^{(0)\mu\beta\nu\alpha}\right)$, $O_{\rm \sym}^{(2)\mu\alpha\nu\beta}=\frac12\left(O^{(2)\mu\alpha\nu\beta}+O^{(2)\mu\beta\nu\alpha}\right)$.
We can write $O_{\rm \sym}^{\mu\alpha\nu\beta}$ as 
\begin{equation}\label{Osdecomp}
O_{\rm \sym}^{\mu\alpha\nu\beta}=O_{\rm \sym}^{(0)\mu\alpha\nu\beta}+O_{\rm \sym}^{(2)\mu\alpha\nu\beta}+O_{\rm \sym}^{(r)\mu\alpha\nu\beta}.
\end{equation}
The last term $O_{\rm \sym}^{(r)\mu\alpha\nu\beta}$ can be expressed explicitly as
\begin{equation}\label{Ors}
O_{\rm \sym}^{(r)\mu\alpha\nu\beta}=-\frac14\left(\epsilon^{\mu\alpha\rho\sigma}\epsilon^{\nu\beta\kappa\lambda}+\epsilon^{\mu\beta\rho\sigma}\epsilon^{\nu\alpha\kappa\lambda}\right)G^{A}_{\rho\kappa}G^{A}_{\sigma\lambda}-\mbox{all possible traces}.
\end{equation}
Since a product of two Levi-Civita tensors can be expressed in terms of the metric tensor,
Eq.~(\ref{Ors}) can be shown to be equivalent to Eq.~(\ref{Osdecomp}). 

The matrix elements of $O_\sym^{\mu\alpha\nu\beta}$ between spin-averaged
proton states can be expressed in terms of  the matrix elements of
$G^{A\mu\alpha}G^{A\nu\beta}$. These can depend only on the metric
tensor and powers of the proton four-momentum, $k$. We can have six
structures $g^{\mu\nu}g^{\alpha\beta}$, $g^{\mu\beta}g^{\nu\alpha}$,
$g^{\mu\nu}k^\alpha k^\beta$, $g^{\alpha\beta}k^\mu k^\nu$,
$g^{\mu\beta}k^\nu k^\alpha$, and $g^{\nu\alpha}k^\mu k^\beta$ with
arbitrary coefficients.%
\footnote{The structure $k^\mu k^\nu
  k^\alpha k^\beta$ is symmetric in $\mu\leftrightarrow \alpha$ and
  $\nu\leftrightarrow \beta$.
}
The matrix elements of
$G^{A\mu\alpha}G^{A\nu\beta}$ are antisymmetric under
$\mu\leftrightarrow \alpha$ and $\nu\leftrightarrow \beta$. As a
result, only two of the six coefficients are  independent. These can
be identified with the matrix elements of the spin-0 and spin-2
gluonic operators. In particular this implies that the matrix elements
of  $O_{\rm \sym}^{(r)\mu\alpha\nu\beta}$ between spin-averaged proton
states vanish. 
Working in $d$-dimensions we find the useful operator relation,
\begin{equation}
\dfrac{q_\alpha q_\beta}{q^2} G^{A\mu\alpha}G^{A\nu\beta}=-\dfrac1{d(d-1)}O_g^{(0)}\left(-g^{\mu\nu}+\dfrac{q^\mu q^\nu}{q^2}\right)+\dfrac{q_\alpha q_\beta}{q^2}O^{(2)\mu\alpha\nu\beta}+\dfrac{q_\alpha q_\beta}{q^2}O_{\rm \sym}^{(r)\mu\alpha\nu\beta},
\end{equation}
from which we obtain the expression for $T_c^{\mu\nu}$.
 
\section{Chiral lagrangian constraints \label{sec:chiral}}

\begin{figure}[htb]
\centering
\includegraphics[height=0.5\textwidth]{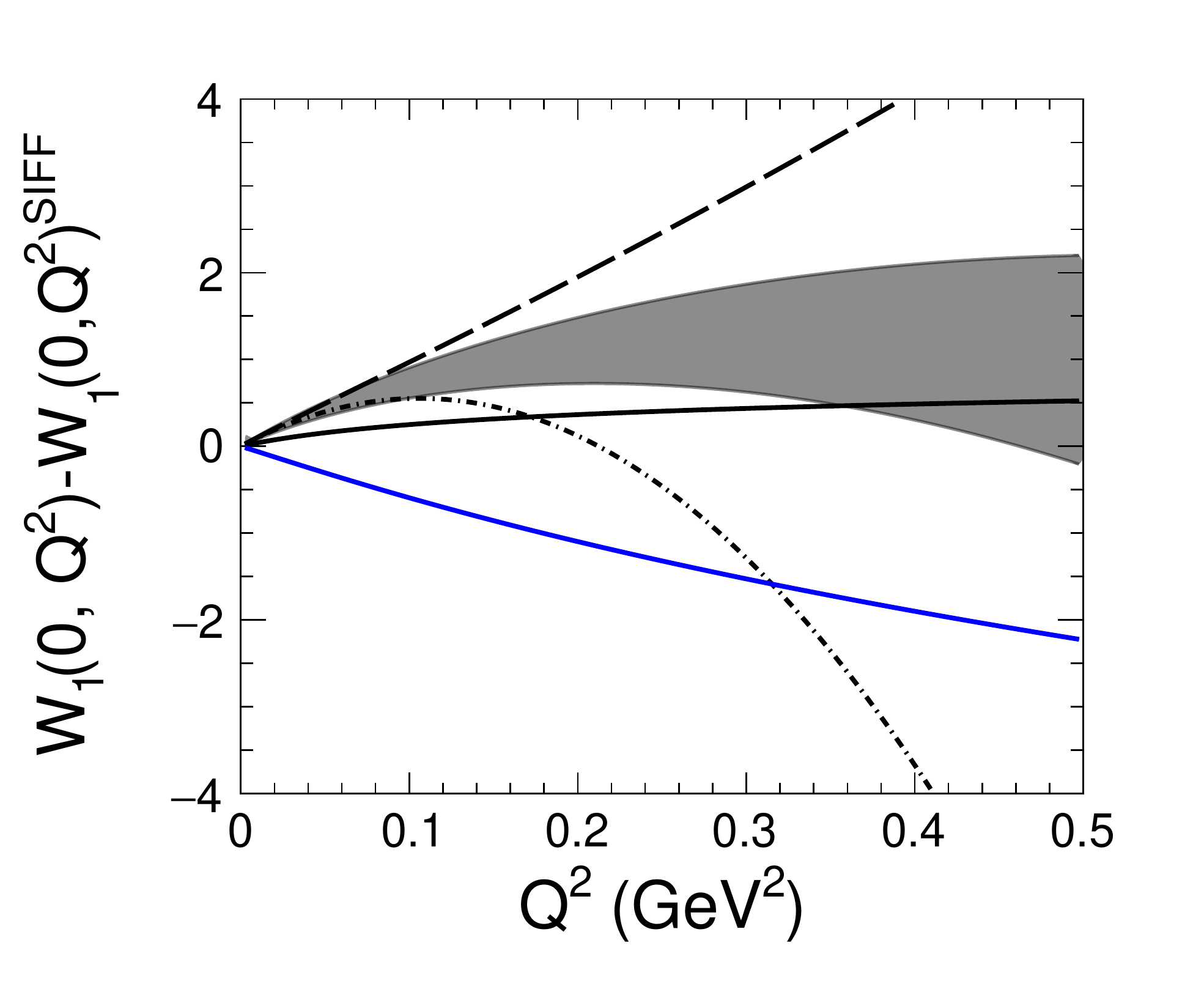}
\caption{Different calculations of $W_1(0,Q^2)-W_1(0,Q^2)^{\rm
    SIFF}$. The black solid line, black dashed line, and solid band,
  give  the ``third-order'' ($\overline T_1^{(3)}$),  ``fourth-order''
  ($\overline T_1^{(3)}+\overline T_1^{(4)}$), and ``fourth-order plus
  $\Delta$'' ($\overline T_1^{(3)}+\overline T_1^{(4)}+\overline
  T_1^{(\Delta)}$) results of Ref.~\cite{Birse:2012eb}, using the
  parameters of Ref.~\cite{Birse:2012eb} and their prescription for
  determining $\delta \beta$. The solid band adds in quadrature the
  errors from $c_1,c_2,c_3,\beta$ and
  $g_{\pi{\scriptscriptstyle\text{N}\Delta}}$.  The dash-dotted black
  line gives the fourth order plus $\Delta$ contribution ($\overline
  T_1^{(3)}+\overline T_1^{(4)}+\overline T_1^{(\Delta)}$), Taylor
  expanded through $\order(Q^4)$.   The bottom blue solid line gives
  the ``third order'' ($T_1^{(\mathrm{NB})}(0,Q^2)$) result of
  Ref.~\cite{Alarcon:2013cba}, using its parameters.\label{fig:W1pol}}
\end{figure}

Let us consider potential improvements to the low-$Q^2$ constraints on the interpolation displayed
in Fig.~\ref{fig:W1_interp}.
Recall that the results (\ref{eq:lowQ2}) are obtained by a straightforward computation
from the NRQED lagrangian.  The relevant NRQED Wilson coefficients are determined
by Taylor expansions of conventionally-defined
relativistic form factors and Compton scattering amplitudes.
In principle, a similar strategy can be applied at higher orders
in the low-$Q^2$ expansion.  However, the NRQED lagrangian is not
available through the relevant ($1/M^5$) order, nor has the
scattering data yet been analyzed with this application in mind.

\begin{figure}[t]
  \begin{center}
    \includegraphics[height=0.5\textwidth]{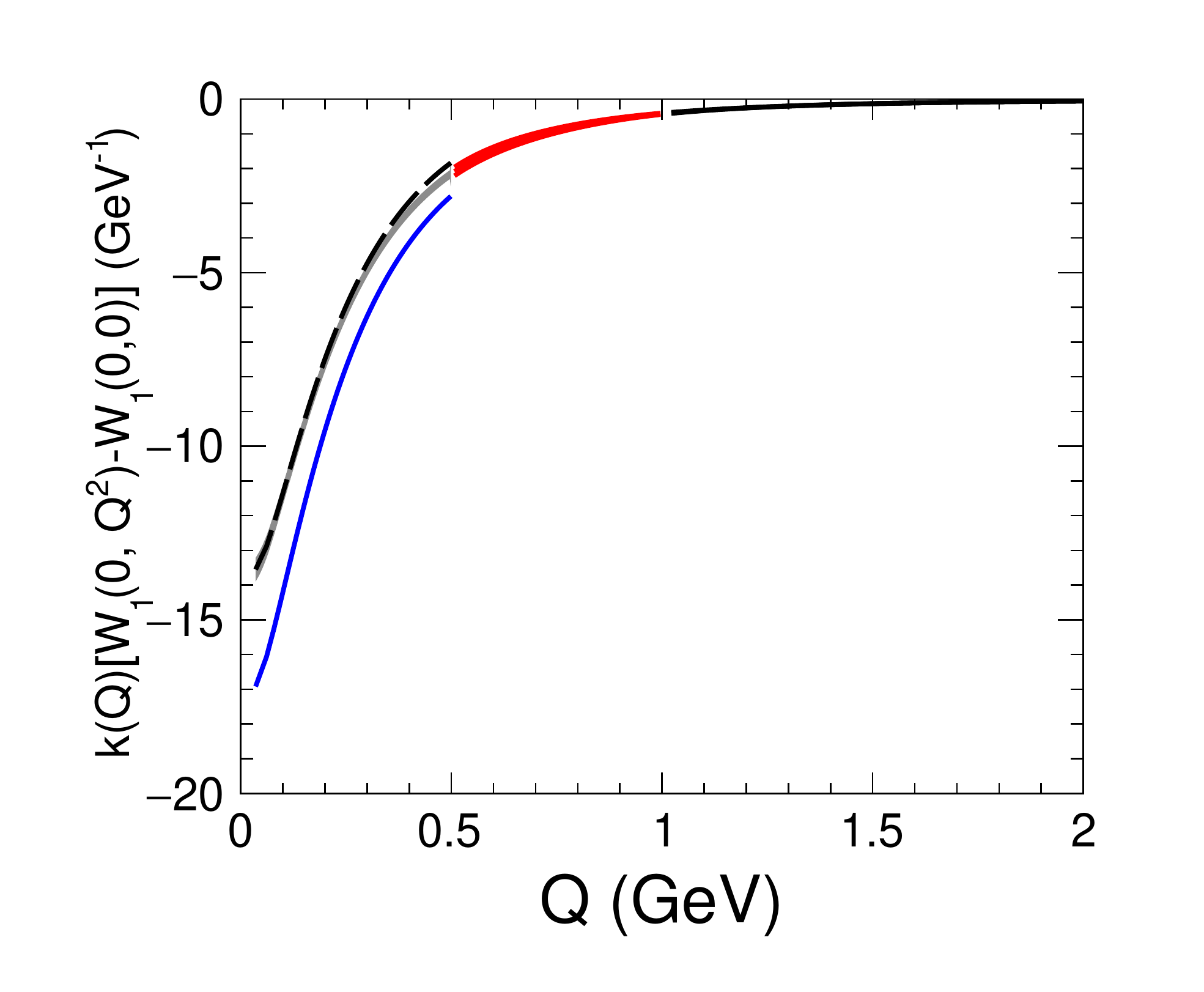}
    \caption{
      Same as Fig.~\ref{fig:W1_moment}, but with 
      $W_1(0,Q^2)$ at low-$Q^2$ taken as $W_1(0,Q^2)^{\rm SIFF}$ with dipole nucleon
      form factors, and deviation from $W_1(0,Q^2)^{\rm SIFF}$ given by the
      band in Fig.~\ref{fig:W1pol}.  
      \label{fig:W1_moment_SIFF}}
  \end{center}
\end{figure}

Chiral lagrangian analysis can in principle be used to further constrain
$W_1(0,Q^2)$ in the regime where a finite number of hadronic excitations are
relevant.%
\footnote{
  We remark that this is a distinct application of chiral lagrangian analysis compared to the
  two-step matching of QCD onto NRQED
  with chiral lagrangian as intermediate theory~\cite{Peset:2014jxa}. 
  We are here directly matching QCD onto NRQED, using experimental data supplemented
  with constraints on $W_1(0,Q^2)$ to determine the forward Compton amplitude as the relevant hadronic
  input.  
}
As usual, the error budget should account for scheme choices, such as
whether $n_f=2$ or $n_f=3$ light flavors are considered,
whether formally higher-order terms are included at each order in the expansion,
and how $\Delta$ excitations are treated. 
Fig.~\ref{fig:W1pol} shows the low-$Q^2$ behavior of
$W_1(0,Q^2) - W_1(0,Q^2)^{\rm SIFF}$ in several calculational schemes,
where $W_1(0,Q^2)^{\rm SIFF}$ denotes a conventional subtraction term
involving elastic form factors.%
\footnote{
  A transcription error appears in
  Ref.~\cite{Birse:2012eb}: the second term in (A.4) should have the opposite sign~\cite{Birse:pc}.
}
Let us consider at face value the error budget from Ref.~\cite{Birse:2012eb}
to constrain the deviation from $W_1(0,Q^2)^{\rm SIFF}$ below $Q^2=(500\,{\rm MeV})^2$.
Neglecting the uncertainty on the SIFF contribution,%
\footnote{Form factor curvature and higher-order terms in the
  Taylor expansion of form factors are poorly constrained by data, but the existence of
  data-driven constraints on $F_i(q^2)$ throughout the relevant range of $q^2$ should
  drive a relatively small uncertainty for this contribution.
}
the corresponding
interpolation is displayed in Fig.~\ref{fig:W1_moment_SIFF}.
Also displayed in Fig.~\ref{fig:W1_moment_SIFF} are upper and lower envelopes corresponding
to alternative schemes displayed in Fig.~\ref{fig:W1pol}. 

Whether it is possible to compute the difference $W_1(0,Q^2)-W_1(0,Q^2)^{\rm SIFF}$
more precisely than $W_1(0,Q^2)$ itself,
and how best to assign an error budget that accounts
for computational scheme dependence in Fig.~\ref{fig:W1pol}, remain difficult
questions.
Figure~\ref{fig:W1_moment_SIFF} illustrates the potential to combine low-$Q^2$
constraints with the OPE. 

\end{appendix}

\end{fmffile}

\end{document}